\documentclass[12pt,aps,nopreprint,superscriptaddress,
longbibliography,amsmath,amssymb,showpacs]{revtex4-2}
\usepackage{xcolor,hyperref}

\hypersetup{colorlinks,linkcolor={blue!50!black},
   urlcolor={blue!80!black}}
\usepackage{graphicx,float}
\usepackage{tabularx}
\usepackage{booktabs}
\usepackage{multirow}
\usepackage{orcidlink}
\usepackage{float}

\begin{document}

    \title{A Non-Equilibrium Dissipation Parameter and the Ideal Glass} 
    \author{Jun-Ying Jiang\orcidlink{0000-0001-6760-3664}} 
        \thanks{These authors contributed equally.}
    \author{Liang Gao\orcidlink{0009-0000-2926-7364}}
        \thanks{These authors contributed equally.}
    \author{Hai-Bin 
    Yu\orcidlink{0000-0003-0645-0187}}\email{haibinyu@hust.edu.cn}
    \affiliation{Wuhan National High Magnetic Field Center and School of Physic, Huazhong University of Science and Technology, Wuhan 430074, Hubei, 
    China}
    \date{\today}

\begin{abstract}
\textbf{Glass materials, as quintessential non-equilibrium systems, exhibit properties such as energy dissipation that are highly sensitive to their preparation histories. A key challenge has been identifying a unified order parameter to rationalize these properties. Here, we demonstrate that a configurational distance metric can effectively collapse energy dissipation data across diverse preparation histories and testing protocols, including varying cooling rates, aging processes, probing times, and the amplitudes of mechanical excitation, as long as the temperature remains above the so-called ideal glass transition (where the extrapolated structural relaxation time diverges). Our results provide a unified description for the non-equilibrium dissipation and suggest that the putative concept of the ideal glass transition is imprinted in material characteristics.}
\end{abstract}

\maketitle

Unlike equilibrium systems, which are stable and time-invariant, non-equilibrium systems exhibit net flows of energy, matter, or information, leading to complex behaviors like aging and memory effects \cite{Diego2024,baity2023,scalliet2019}. Glass is a classic example of a non-equilibrium system. It forms when a liquid is cooled rapidly, preventing it from crystallizing into an ordered, equilibrium state. Instead, atoms or molecules become ``frozen" in a disordered, amorphous structure, creating a solid that retains some liquid-like properties \cite{V.B2020,Hajime2019,ishino2025a,gao2025,frey2025,Ludovic2011,dyre2024,lunkenheimer2023}. The defining feature of glass materials is their thermodynamic non-equilibrium state, which makes them highly sensitive to their preparation history. Variations in cooling rates \cite{granata2014, atila2025}, annealing times \cite{schawe2022, da2024}, pressure conditions \cite{spie2022, bohmer2024, saini2024}, or aging treatments \cite{Martin2018, Guan2018, peter2009} can significantly influence the materials' microscopic structures and macroscopic properties. For example, rapid cooling leads to a higher-energy, more disordered structure, whereas slow cooling allows the material to approach a more stable, lower-energy state \cite{schawe2019a,greer2015}. Not only does this sensitivity to the preparation history present challenges for the study and application of glass materials, but it also creates opportunities to tune their properties \cite{V.B2020,Ludovic2011,chengnc2022,lazaro2019}.

Although phenomenological models and the ``materials time" concept exist, no unified order parameter has been established to characterize the history-dependent properties of amorphous materials. Recently, an order parameter-the inherent structural minimum displacement (IS $D_{\text{min}}$, Eq. \ref{eq_ISD}) \cite{yunsr2024}, was proposed to theoretically rationalize the underlying mechanism of relaxation kinetics of various amorphous material systems at equilibrium liquid states.

\begin{equation}\label{eq_ISD}
    \text{IS}\,D_{\text{min}}(t)= [\frac{1}{N}\text{min}\sum\limits_{i,j}^{N,N} C_{ij}(t)X_{ij}(t)]^{1/2}
\end{equation}

In Eq.\ref{eq_ISD}, $C_{i,j}(t)=\left[\vec{r_i}(t+t_0)-\vec{r_j}(t_0)\right]^2$ is denoted as the \textit{cost matrix}, with $\vec{r_i}$ is the position of $ i- $th atom, and $ N $ is the number of atoms; $X$ is a $ N \times N $ boolean matrix, where $X[i,j]=1$ if row $i$ is assigned to column $j$ for the minimization of Eq. \ref{eq_ISD} and otherwise $X[i,j] = 0$.

This parameter treats all atoms in the system as a whole, assuming exchange 
symmetry among atoms. The configuration is considered unchanged if atomic 
position-replacing motions occur. It reduces to the root mean square 
displacement \cite{zhouacta2023} when $X$ is a unitary matrix under the 
condition of small displacements (no larger than half interatomic distance). 
When coarse-graining is applied, it becomes the overlap function of Parisi 
\cite{parisi1983,gui2020}. The solution of Eq.\ref{eq_ISD} has been obtained by 
the Hungarian algorithm \cite{crouse2016b}. 

Previous work has shown a power-law scaling relationship between the relaxation 
dissipation of disordered materials (characterized by internal dissipation 
$\delta$) and IS $D_{\text{min}}$, indicating that the relaxation dissipation 
of liquids can be revealed across different materials and time scales. However, 
these findings were limited to steady-state liquids and did not explore its 
applicability to non-equilibrium glassy states \cite{yunsr2024}. Extending this 
parameter to non-equilibrium regimes could provide a unified framework to 
characterize aging, memory, and preparation-history-dependent effects in 
glasses. 

\begin{figure}
	\begin{center}
		\includegraphics[width=0.85\textwidth]{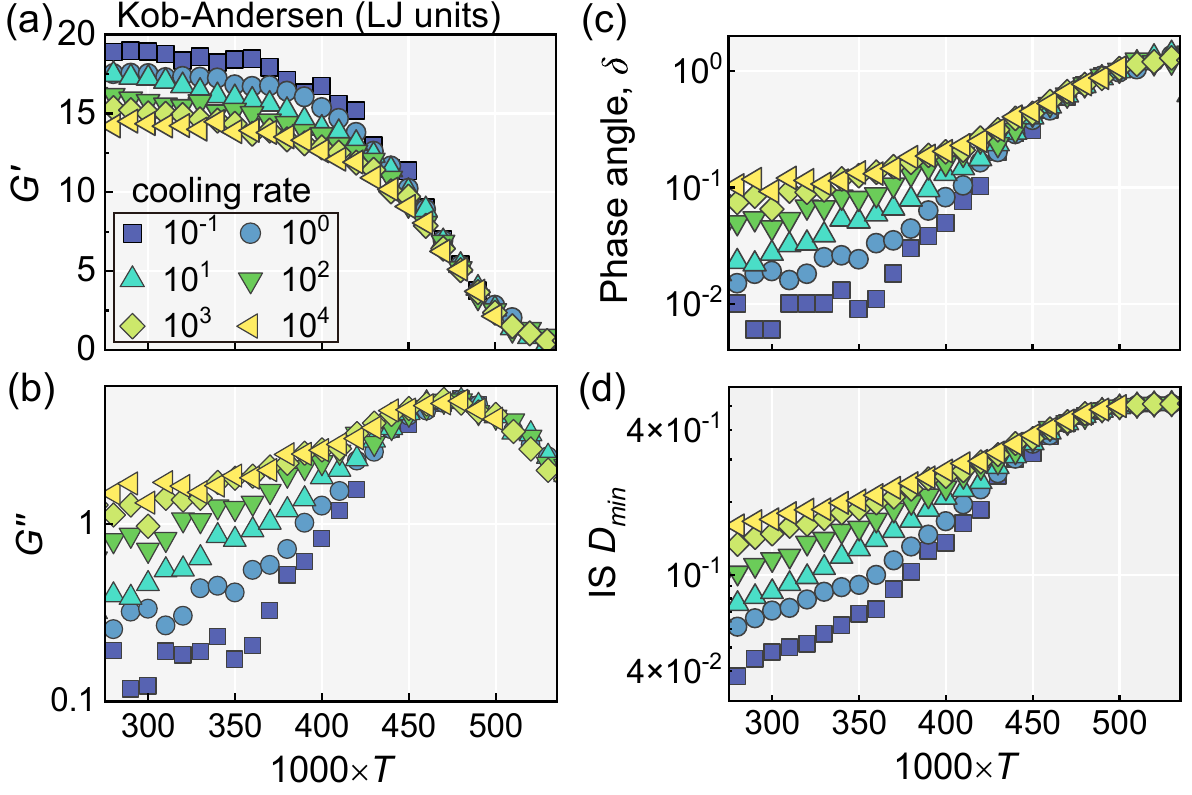}
		\caption{\label{fig:dms}
			Dynamic mechanical response of the Kob-Andersen model obtained from MD-DMS simulations at different cooling rates. 
			(a) for Storage shear modulus ${G}'$, 
			(b) for loss shear modulus ${G}''$, 
			(c) for phase angle $\delta$, 
			and (d) for IS $D_{\text{min}}$. 
		}
	\end{center}
\end{figure}

Moreover, the non-equilibrium glassy state invites a long-standing question 
about the so-called ideal glass transition and ideal glass state. The ideal 
glass 
transition refers to a pure thermodynamic ideal phase transition that occurs 
when a supercooled liquid is cooled to the Kauzmann temperature ($T_K$), where 
the 
extrapolated entropy of the liquid equals that of the crystal, 
eliminating kinetic effects. In many cases, the Kauzmann temperature is 
demonstrated to be the temperature at which the viscosity diverges ($T_0$), 
i.e., $T_K = T_0$ \cite{Tanaka2003}.

The concept of ideal glass transition is considered as a rescue of the 3rd 
thermodynamic law in liquid science. However, as the liquid becomes very slow 
in dynamics, no real experiments can probe it. There have been substantial 
discussions and debates around it \cite{dyre2008, Adhikari2023}. Whether such a 
thermodynamic 
limit exists, and whether it can be captured by an appropriate structural or 
dynamical order parameter, remains one of the key challenges in glass physics.

Several attempts have been conducted to verify the existence of $T_0$ or the ideal glass state \cite{yoon2018, Monnier2021,cammarota2012}. For example, Ozawa et al. \cite{ozawa2018a} conducted a computer simulation study on the thermodynamic and kinetic characteristics of the glass morphology that undergoes ideal glass transition due to randomly fixed particles. They discovered that even in the state of deep equilibrium, the particles would explore multiple inherent structures (IS) within the local minimum of free energy. 

\begin{figure}[h]
	\begin{center}
		\includegraphics[width=0.8\textwidth]{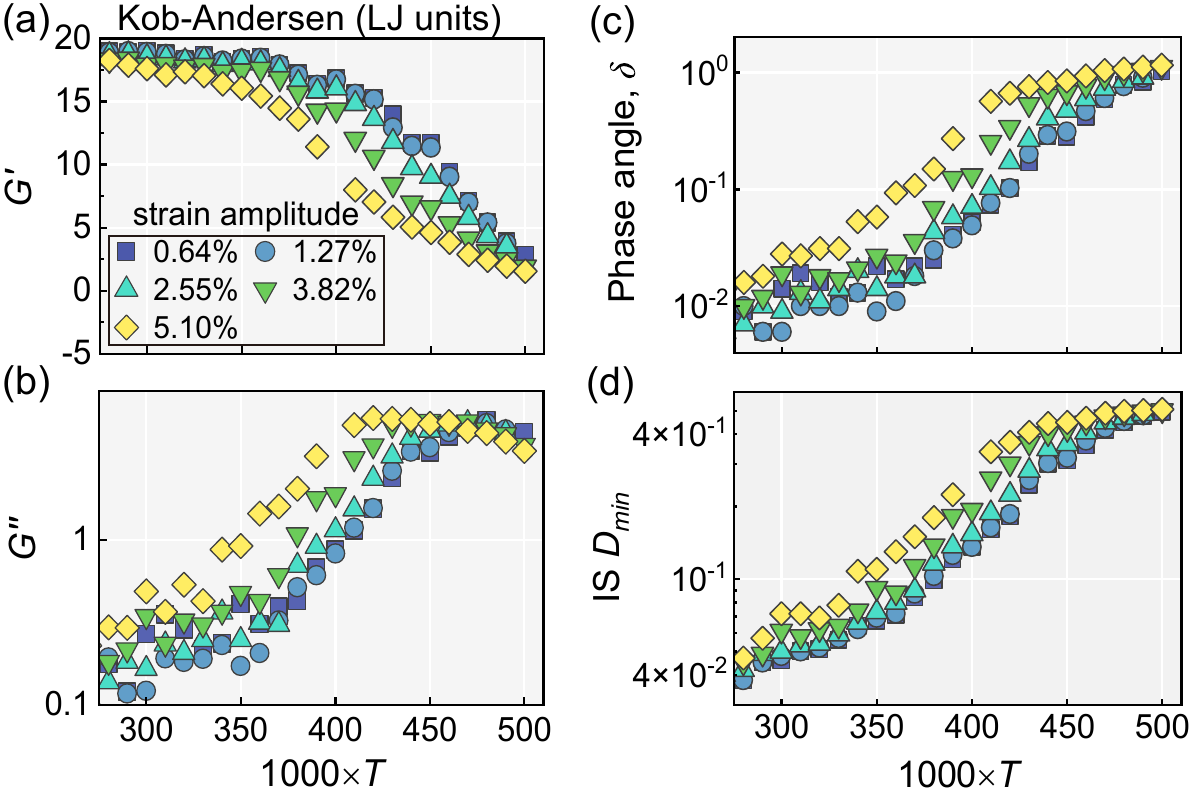}
		\caption{\label{fig:amp}
			Strain amplitude influence on dynamics of Kob-Andersen model.
			(a) for Storage shear modulus ${G}'$, 
			(b) for loss shear modulus ${G}''$,
			(c) for phase angle $\delta$, and
			(d) for IS $D_{\text{min}}$. This model glass is cooled with a cooling rate of 0.1 in LJ unit. 
		}
	\end{center}
\end{figure}

So far, no unified order parameter has been proposed to describe its 
characteristics. The concept of local IS clusters within minima aligns directly 
with the structural metric IS $D_{\text{min}}$, which quantifies inherent 
differences in precisely such local configuration basins. IS $D_{\text{min}}$ 
thus offers a natural framework to characterize the structural heterogeneity 
governing partial relaxation near arrest. 

This study aims to validate whether the order parameter IS $D_{\text{min}}$ can describe the relaxation-dissipation behavior of non-equilibrium glasses and what happens around $T_0$. Our results show that IS $D_{\text{min}}$ is related to the loss phase angle when the temperature is kept above the ideal glass transition, including different cooling rates, aging processes, probing times, and mechanical excitations. In contrast, below the ideal glass transition temperature, IS $D_{\text{min}}$ decouples from the phase angle without direct correlation. These findings not only provide a much-needed order parameter for characterizing non-equilibrium dissipation but also lend substantial support to the concept of ideal glass.

As a typical example, we study the Kob-Andersen binary Lennard-Jones (LJ) mixture, consisting 80\% large A particles and 20\% small B particles \cite{Kob1993, Andersen2008}. In this mixture, all particles have the same mass and interact via LJ pair potentials $v(r) = 4\varepsilon\left[(r/\sigma)^{-12} - (r/\sigma)^{-6}\right]$, truncated and shifted to zero at $2.5\sigma$. The interaction parameters are $\sigma_{BB}/\sigma_{AA} = 0.88$, $\sigma_{AB}/\sigma_{AA} = 0.8$, $\varepsilon_{BB}/\varepsilon_{AA} = 0.5$, and $\varepsilon_{AB}/\varepsilon_{AA} = 1.5$. All quantities are in the LJ units: length in units of $\sigma_{AA}$, temperature in units of $\sigma_{AA}/k_{\text{B}}$, time in the units of $\sqrt{m\sigma^2_{AA}/\varepsilon_{AA}}$, pressure and modulus in the units of $\sigma^3_{AA}/ \varepsilon_{AA}$. We kept the number of density $N/V=1.2$. This model has good glass-forming ability and enables us to study equilibrium supercooled liquids, and glasses with cooling rates ranging from $10^4$ to $10^{-1}$ in LJ units (the functional relationship between temperature and potential energy is shown in FIG.~S1). 

Using molecular dynamics simulations of dynamic mechanical spectra (MD-DMS) in LAMMPS \cite{lmp22, lmp13}, we explore the relaxation-dissipation behaviors in both equilibrium liquids and non-equilibrium glasses \cite{lyu2021, yu2017, Takeshi2022,Douglas2024}. In MD-DMS, a sinusoidal shear strain is applied with an oscillation period $t_\omega$ (related to frequency $f=1/t_\omega$) and a strain amplitude $\varepsilon_A$. The resulting stress $\sigma(t)$ and the phase difference $\delta$ between stress and strain are measured and adapt by $ \sigma (t) = \sigma_0 + \sigma_{A} \text{sin} ( 2 \pi t / t_{\omega} + \delta) $. The storage and loss moduli are calculated by $G' = \sigma_A / \varepsilon_A \text{cos} (\delta)$ and $G'' = \sigma_A / \varepsilon_A \text{sin} (\delta)$, respectively. 

We determine the $\alpha$ relaxation time $\tau_{\alpha}$ and the characteristic period $t_{w}$ from the peak time of the isothermal and equilibrium MD-DMS (see FIG.~S2). Afterwards, these data can be well fitted by a Vogel-Fulcher-Tammann (VFT) function \cite{ngai2023,Guan2018}, $\tau_{\alpha} = \tau_0 \exp(DT_0/(T-T_0))$, with a divergent temperature $T_0 = 0.27$ (FIG.~S3). This temperature has been considered the ideal glass transition temperature; below which, if the liquid could be equilibrated, it would enter an ideal glass phase.

\begin{figure}[h]
\begin{center}
    \includegraphics[width=0.5\textwidth]{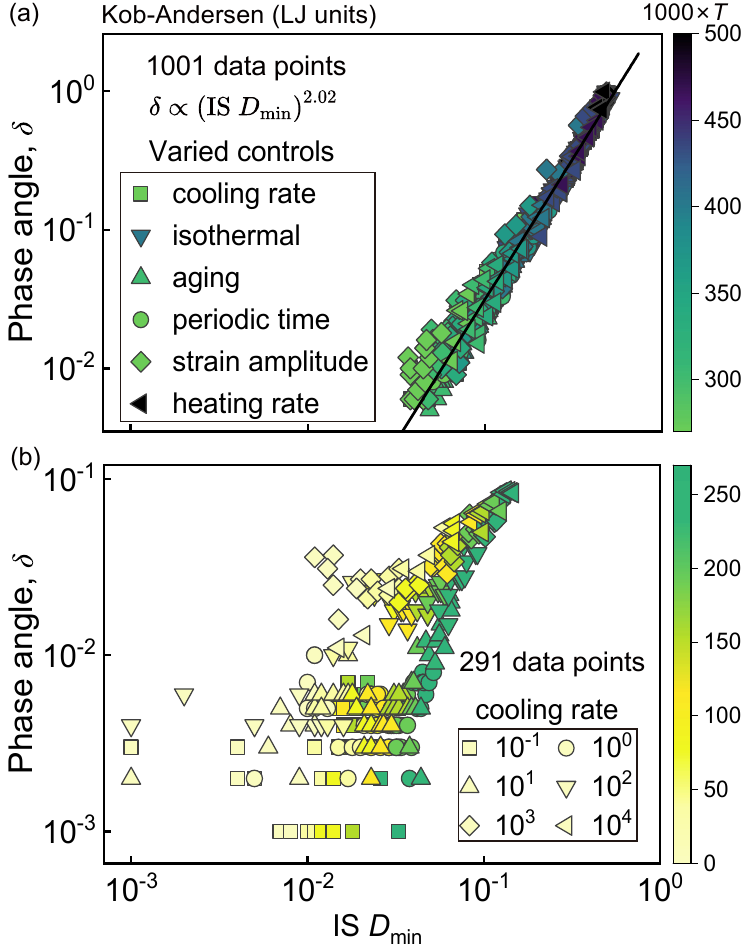}
    \caption{\label{fig:3}
    Correlation between phase angle $\delta$ and IS $D_{\text{min}}$ in the Kob–Andersen model at temperatures (a) above and (b) below $T_0$ ($\sim$ 0.27). 
    The legend entries correspond to different simulation protocols: 
    `cooling rate' represents samples prepared under different cooling rates (FIG.~\ref{fig:dms}); 
    `isothermal' refers to samples equilibrated at fixed temperatures (FIG.~S4); 
    `aging' indicates samples aged at various temperatures after quenching (FIG.~S5); 
    `periodic time' denotes the various time intervals in MD-DMS at the slowest cooling rate (FIG. S2); 
    `strain amplitude' refers to different amplitudes of applied strain in MD-DMS (FIG.~\ref{fig:amp}); and 
    `heating rate' represents samples subjected to varying heating rates (FIG.~S6). 
     Comparative results of $\delta$ and IS $D_{\text{min}}$ across all conditions are shown in FIG. S7.
    }
\end{center}
\end{figure}

Figure~\ref{fig:dms} shows typical relaxation dynamics data for varied cooling rates, with a single probing time of $t_w = 1000$ in LJ time unit. At temperatures below $0.45$, the system exhibits characteristic non-equilibrium behaviors. With decreasing cooling rate, the storage modulus $G'$ increases, while the loss modulus $G''$ and the phase angle $\delta$ decrease. Specifically, the variation of $\delta$ can be up to 10 times at lower temperatures, particularly for $T < 0.35$. In FIG. \ref{fig:dms}(d), we also present the IS $D_{\text{min}}$ during the MD-DMS. Similarly as in (c), these values exhibit a clear dependence on the cooling rate, with slower cooling resulting in smaller $D_{\text{min}}$.

In addition, we have explored non-equilibrium dynamics under various conditions, including strain amplitude (FIG. \ref{fig:amp}), equilibrium isothermal (by varying the probing $t_w$, FIG.~S4), aging (FIG.~S5), and heating rate (FIG.~S6). For example, as shown in FIG.~\ref{fig:amp}, increasing the strain amplitude for MD-DMS leads to higher values of both $\delta$ and IS $D_{\text{min}}$, and shifts the $\alpha$-relaxation to lower temperatures. These observations highlight characteristic non-linear and non-equilibrium behaviors.

Figures~\ref{fig:3}(a) and~\ref{fig:3}(b) plot $\delta$ against IS $D_{\text{min}}$ in the Kob–Andersen model for temperatures above and below $T_0$, respectively. We find, in FIG.~\ref{fig:3}(a), the data $T > T_0$ are well correlated. A power-law fit using Eq.~\ref{eq_scaling} yields an exponent of $b = 2.02 \pm 0.03$.
\begin{equation}\label{eq_scaling}
\delta \propto \left( \text{IS} \ D_{\text{min}} \right)^b
\end{equation}
These results illustrate that the non-equilibrium dissipation at $T>T_0$ can be uniquely determined by IS $D_{\text{min}}$, consistent with our previous findings on equilibrium dynamics. Remarkably, we now find that all the dissipation data, including those from non-equilibrium glasses, equilibrium supercooled liquids, and strain-driven non-linear data, follow this relation.

\begin{figure}[h]
	\begin{center}
		\includegraphics[width=1\textwidth]{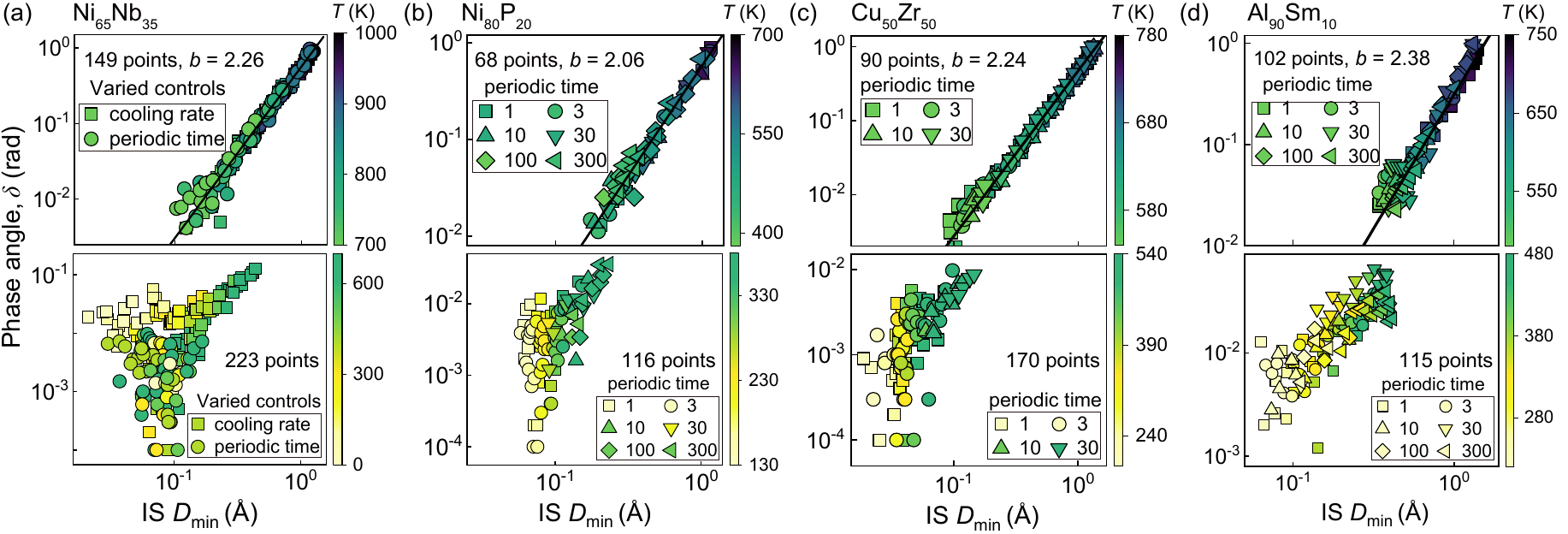}
		\caption{\label{fig:4}
			Identical correlation between the phase $\delta$ (rad) and IS $D_{\text{min}}$ (\AA) for 
			(a) Ni$_{65}$Nb$_{35}$, 
			(b) Ni$_{80}$P$_{20}$, 
			(c) Cu$_{50}$Zr$_{50}$ and 
			(d) Al$_{90}$Sm$_{10}$ models, which have $T_0$ of 490 K, 390 K, 540 K, and 480 K, respectively.
			The corresponding details are provided in FIGs.~S8-S14, respectively. 
		}
	\end{center}
\end{figure}

On the other hand, the data are scattered at $T < T_0$ in FIG.~\ref{fig:3}(b), and $\delta$ and IS $D_{\text{min}}$ are no longer correlated in a united manner. It is intriguing that $T_0$ defines the temperature range where Eq.~\ref{eq_scaling} holds. This implies that $T_0$ might indeed be a relevant temperature for non-equilibrium glass states. 

Our results are not limited to the Kob-Andersen models; we have also verified that the main findings apply to metallic glass models with many body interactions. As shown in Figure~\ref{fig:4}, the same conclusions as in Figure~\ref{fig:3} are obtained for the (a) Ni$_{65}$Nb$_{35}$ \cite{Zhang16-NiNb}, (b) Ni$_{80}$P$_{20}$, (c) Cu$_{50}$Zr$_{50}$, and (d) Al$_{90}$Sm$_{10}$ models (data for these models were sourced from \cite{gao2025}). Additional details on these metallic glass models are provided in \textit{Supplementary Material}.

\begin{figure}[h]
	\begin{center}
		\includegraphics[width=0.6\textwidth]{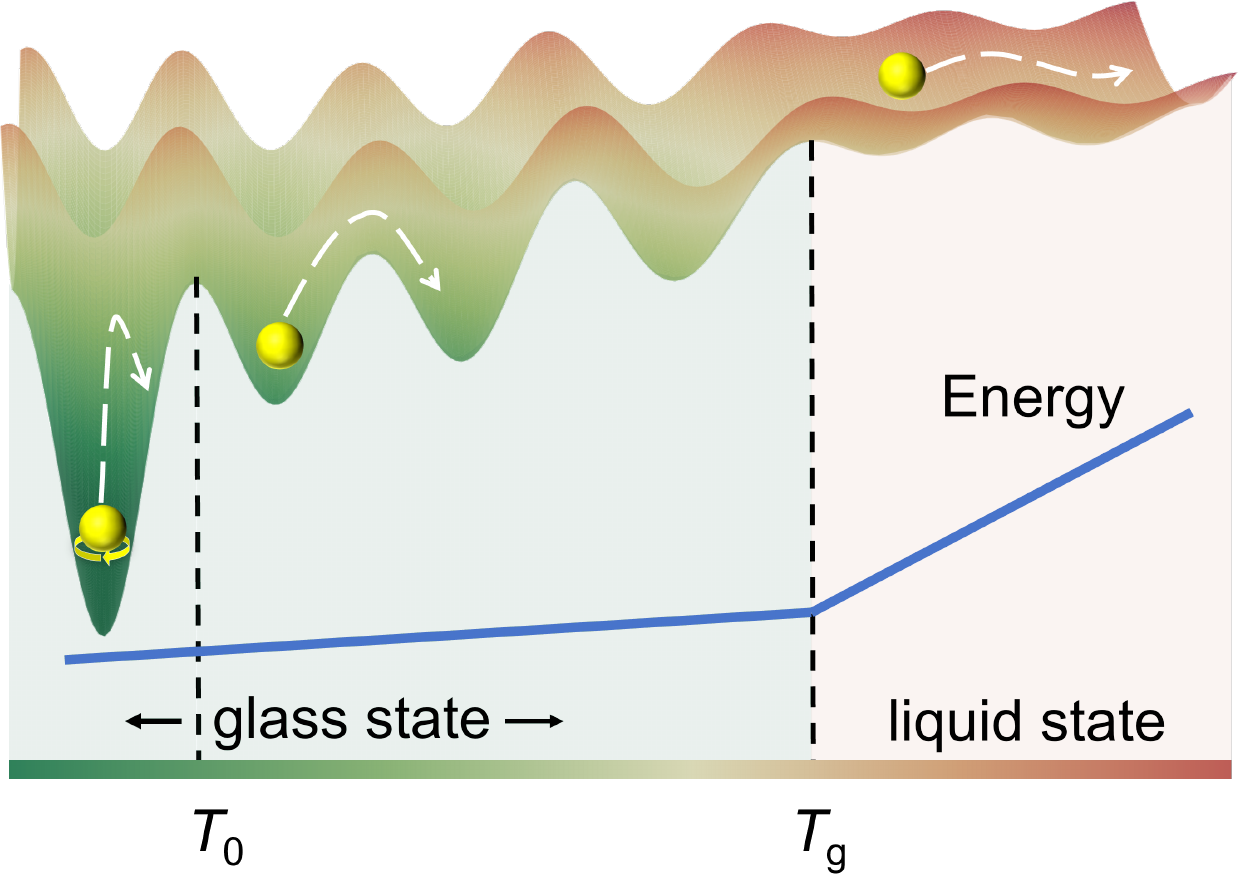}
		\caption{\label{fig:model}
			A schematic of potential energy landscape (PEL) illustrates the process of dissipation relaxation. At temperatures above the $T_g$, the structural energy barriers are sufficiently low, allowing the system to undergo frequent structural transformations. In the intermediate temperature range, $T_0 < T < T_g$, the system can across barriers. When the temperature falls below $T_0$, the energy barriers become so high that structural rearrangements are essentially suppressed.
		}
	\end{center}
\end{figure}

Why is the order parameter IS $D_{\text{min}}$ effective for describing non-equilibrium dissipation when $T > T_0$? One plausible explanation involves the potential energy landscape (PEL) \cite{pel1,dingnc2016, Takeshi2024,fan2017}, which is a potential energy function of an $N$-body system, $\varPhi(r_1, r_2, \ldots, r_N)$. Here, the vectors $r_i$ include position, orientation, and intermolecular coordinates. For example, for a system of $N$ identical atoms, this landscape is a $(3N + 1)$-dimensional object. As schematically shown in FIG.~\ref{fig:model}, IS $D_{\text{min}}$ characterizes the shortest distance between two local energy minima (inherent structures).

Previous work has shown that the power-law scaling in Eq.~\ref{eq_scaling} results from activation between different local minima with distinct configurations \cite{yunsr2024}. Given that energy dissipation is associated with activation, and assuming the energy barrier is harmonic, one obtains $\delta \propto \left( \text{IS} \ D_{\text{min}} \right)^2$. However, Eq.~\ref{eq_scaling} with a power $b > 2$ suggests that the activation barrier is not purely harmonic. Instead, $b$ characterizes the local curvature of the PEL.

The breakdown of Eq.~\ref{eq_scaling} at $T < T_0$ may be due to high energy barriers and insufficient thermal energy for activation. This suggests that dissipation at $T < T_0$ is primarily due to other motions rather than the configurational changes captured by IS $D_{\text{min}}$. Atomic vibrations are one such motion, quasi-localized vibrations are dissipative but would not change the IS $D_{\text{min}}$. For example, the so-called Boson peak could yield substantial dissipation in the low temperature range \cite{haraL2025a}.  Additionally, liquid-like clusters that are dissipative but non-diffusive have recently been identified as the source of low-temperature $\gamma$ relaxation in glasses. These arguments imply that relaxation dissipation at low temperatures is different from the normal glass ($T>T_0$). These arguments are consistent with the prevalent view that the glass have a PEL dominated temperature range.

To summarize, this work proposes the order parameter IS $D_{\text{min}}$ as a universal descriptor of non-equilibrium dissipation in glassy materials. Specifically, the ideal glass transition temperature $T_0$ emerges as a fundamental threshold: dissipation follows a power-law scaling with IS $D_{\text{min}}$ for $T > T_0$ through activation over PEL barriers; while below $T_0$, this relationship breaks down as dissipation shifts to non-configurational mechanisms. Our findings provide a fundamental framework for understanding and predicting dissipation mechanisms in glassy materials.

\begin{acknowledgments}
The computational work was carried out on the public computing service platform provided by the Network and Computing Center of HUST. We are thankful for the support of the National Science Foundation of China (52071147).
\end{acknowledgments}


\begin{thebibliography}{52}%
	\makeatletter
	\providecommand \@ifxundefined [1]{%
		\@ifx{#1\undefined}
	}%
	\providecommand \@ifnum [1]{%
		\ifnum #1\expandafter \@firstoftwo
		\else \expandafter \@secondoftwo
		\fi
	}%
	\providecommand \@ifx [1]{%
		\ifx #1\expandafter \@firstoftwo
		\else \expandafter \@secondoftwo
		\fi
	}%
	\providecommand \natexlab [1]{#1}%
	\providecommand \enquote  [1]{``#1''}%
	\providecommand \bibnamefont  [1]{#1}%
	\providecommand \bibfnamefont [1]{#1}%
	\providecommand \citenamefont [1]{#1}%
	\providecommand \href@noop [0]{\@secondoftwo}%
	\providecommand \href [0]{\begingroup \@sanitize@url \@href}%
	\providecommand \@href[1]{\@@startlink{#1}\@@href}%
	\providecommand \@@href[1]{\endgroup#1\@@endlink}%
	\providecommand \@sanitize@url [0]{\catcode `\\12\catcode `\$12\catcode
		`\&12\catcode `\#12\catcode `\^12\catcode `\_12\catcode `\%12\relax}%
	\providecommand \@@startlink[1]{}%
	\providecommand \@@endlink[0]{}%
	\providecommand \url  [0]{\begingroup\@sanitize@url \@url }%
	\providecommand \@url [1]{\endgroup\@href {#1}{\urlprefix }}%
	\providecommand \urlprefix  [0]{URL }%
	\providecommand \Eprint [0]{\href }%
	\providecommand \doibase [0]{https://doi.org/}%
	\providecommand \selectlanguage [0]{\@gobble}%
	\providecommand \bibinfo  [0]{\@secondoftwo}%
	\providecommand \bibfield  [0]{\@secondoftwo}%
	\providecommand \translation [1]{[#1]}%
	\providecommand \BibitemOpen [0]{}%
	\providecommand \bibitemStop [0]{}%
	\providecommand \bibitemNoStop [0]{.\EOS\space}%
	\providecommand \EOS [0]{\spacefactor3000\relax}%
	\providecommand \BibitemShut  [1]{\csname bibitem#1\endcsname}%
	\let\auto@bib@innerbib\@empty
	\bibitem [{\citenamefont {Tapias}\ \emph {et~al.}(2024)\citenamefont {Tapias},
		\citenamefont {Marteau}, \citenamefont {Aguirre-L\'opez},\ and\ \citenamefont
		{Sollich}}]{Diego2024}%
	\BibitemOpen
	\bibfield  {author} {\bibinfo {author} {\bibfnamefont {D.}~\bibnamefont
			{Tapias}}, \bibinfo {author} {\bibfnamefont {C.}~\bibnamefont {Marteau}},
		\bibinfo {author} {\bibfnamefont {F.}~\bibnamefont {Aguirre-L\'opez}},\ and\
		\bibinfo {author} {\bibfnamefont {P.}~\bibnamefont {Sollich}},\ }\bibfield
	{title} {\bibinfo {title} {Bringing together two paradigms of nonequilibrium:
			Fragile versus robust aging in driven glassy systems},\ }\href
	{https://doi.org/10.1103/PhysRevLett.133.197101} {\bibfield  {journal}
		{\bibinfo  {journal} {Phys. Rev. Lett.}\ }\textbf {\bibinfo {volume} {133}},\
		\bibinfo {pages} {197101} (\bibinfo {year} {2024})}\BibitemShut {NoStop}%
	\bibitem [{\citenamefont {{Baity-Jesi}}\ \emph {et~al.}(2023)\citenamefont
		{{Baity-Jesi}}, \citenamefont {Calore}, \citenamefont {Cruz}, \citenamefont
		{Fernandez}, \citenamefont {{Gil-Narvion}}, \citenamefont {{Gonzalez-Adalid
				Pemartin}}, \citenamefont {{Gordillo-Guerrero}}, \citenamefont {I{\~n}iguez},
		\citenamefont {Maiorano}, \citenamefont {Marinari}, \citenamefont
		{{Martin-Mayor}}, \citenamefont {{Moreno-Gordo}}, \citenamefont
		{Mu{\~n}oz~Sudupe}, \citenamefont {Navarro}, \citenamefont {Paga},
		\citenamefont {Parisi}, \citenamefont {{Perez-Gaviro}}, \citenamefont
		{{Ricci-Tersenghi}}, \citenamefont {{Ruiz-Lorenzo}}, \citenamefont
		{Schifano}, \citenamefont {Seoane}, \citenamefont {Tarancon},\ and\
		\citenamefont {Yllanes}}]{baity2023}%
	\BibitemOpen
	\bibfield  {author} {\bibinfo {author} {\bibfnamefont {M.}~\bibnamefont
			{{Baity-Jesi}}}, \bibinfo {author} {\bibfnamefont {E.}~\bibnamefont
			{Calore}}, \bibinfo {author} {\bibfnamefont {A.}~\bibnamefont {Cruz}},
		\bibinfo {author} {\bibfnamefont {L.~A.}\ \bibnamefont {Fernandez}}, \bibinfo
		{author} {\bibfnamefont {J.~M.}\ \bibnamefont {{Gil-Narvion}}}, \bibinfo
		{author} {\bibfnamefont {I.}~\bibnamefont {{Gonzalez-Adalid Pemartin}}},
		\bibinfo {author} {\bibfnamefont {A.}~\bibnamefont {{Gordillo-Guerrero}}},
		\bibinfo {author} {\bibfnamefont {D.}~\bibnamefont {I{\~n}iguez}}, \bibinfo
		{author} {\bibfnamefont {A.}~\bibnamefont {Maiorano}}, \bibinfo {author}
		{\bibfnamefont {E.}~\bibnamefont {Marinari}}, \bibinfo {author}
		{\bibfnamefont {V.}~\bibnamefont {{Martin-Mayor}}}, \bibinfo {author}
		{\bibfnamefont {J.}~\bibnamefont {{Moreno-Gordo}}}, \bibinfo {author}
		{\bibfnamefont {A.}~\bibnamefont {Mu{\~n}oz~Sudupe}}, \bibinfo {author}
		{\bibfnamefont {D.}~\bibnamefont {Navarro}}, \bibinfo {author} {\bibfnamefont
			{I.}~\bibnamefont {Paga}}, \bibinfo {author} {\bibfnamefont {G.}~\bibnamefont
			{Parisi}}, \bibinfo {author} {\bibfnamefont {S.}~\bibnamefont
			{{Perez-Gaviro}}}, \bibinfo {author} {\bibfnamefont {F.}~\bibnamefont
			{{Ricci-Tersenghi}}}, \bibinfo {author} {\bibfnamefont {J.~J.}\ \bibnamefont
			{{Ruiz-Lorenzo}}}, \bibinfo {author} {\bibfnamefont {S.~F.}\ \bibnamefont
			{Schifano}}, \bibinfo {author} {\bibfnamefont {B.}~\bibnamefont {Seoane}},
		\bibinfo {author} {\bibfnamefont {A.}~\bibnamefont {Tarancon}},\ and\
		\bibinfo {author} {\bibfnamefont {D.}~\bibnamefont {Yllanes}},\ }\bibfield
	{title} {\bibinfo {title} {Memory and rejuvenation effects in spin glasses
			are governed by more than one length scale},\ }\href
	{https://doi.org/10.1038/s41567-023-02014-6} {\bibfield  {journal} {\bibinfo
			{journal} {Nat. Phys.}\ }\textbf {\bibinfo {volume} {19}},\ \bibinfo {pages}
		{978} (\bibinfo {year} {2023})}\BibitemShut {NoStop}%
	\bibitem [{\citenamefont {Scalliet}\ and\ \citenamefont
		{Berthier}(2019)}]{scalliet2019}%
	\BibitemOpen
	\bibfield  {author} {\bibinfo {author} {\bibfnamefont {C.}~\bibnamefont
			{Scalliet}}\ and\ \bibinfo {author} {\bibfnamefont {L.}~\bibnamefont
			{Berthier}},\ }\bibfield  {title} {\bibinfo {title} {Rejuvenation and memory
			effects in a structural glass},\ }\href
	{https://doi.org/10.1103/PhysRevLett.122.255502} {\bibfield  {journal}
		{\bibinfo  {journal} {Phys. Rev. Lett.}\ }\textbf {\bibinfo {volume} {122}},\
		\bibinfo {pages} {255502} (\bibinfo {year} {2019})}\BibitemShut {NoStop}%
	\bibitem [{\citenamefont {Bapst}\ \emph {et~al.}(2020)\citenamefont {Bapst},
		\citenamefont {Keck}, \citenamefont {Grabska-Barwińska}, \citenamefont
		{Donner}, \citenamefont {Cubuk}, \citenamefont {Schoenholz}, \citenamefont
		{Obika}, \citenamefont {Nelson}, \citenamefont {Back}, \citenamefont
		{Hassabis},\ and\ \citenamefont {Kohli}}]{V.B2020}%
	\BibitemOpen
	\bibfield  {author} {\bibinfo {author} {\bibfnamefont {V.}~\bibnamefont
			{Bapst}}, \bibinfo {author} {\bibfnamefont {T.}~\bibnamefont {Keck}},
		\bibinfo {author} {\bibfnamefont {A.}~\bibnamefont {Grabska-Barwińska}},
		\bibinfo {author} {\bibfnamefont {C.}~\bibnamefont {Donner}}, \bibinfo
		{author} {\bibfnamefont {E.~D.}\ \bibnamefont {Cubuk}}, \bibinfo {author}
		{\bibfnamefont {S.~S.}\ \bibnamefont {Schoenholz}}, \bibinfo {author}
		{\bibfnamefont {A.}~\bibnamefont {Obika}}, \bibinfo {author} {\bibfnamefont
			{A.~W.~R.}\ \bibnamefont {Nelson}}, \bibinfo {author} {\bibfnamefont
			{T.}~\bibnamefont {Back}}, \bibinfo {author} {\bibfnamefont {D.}~\bibnamefont
			{Hassabis}},\ and\ \bibinfo {author} {\bibfnamefont {P.}~\bibnamefont
			{Kohli}},\ }\bibfield  {title} {\bibinfo {title} {Unveiling the predictive
			power of static structure in glassy systems},\ }\href
	{https://doi.org/10.1038/s41567-020-0842-8} {\bibfield  {journal} {\bibinfo
			{journal} {Nat. Phys.}\ }\textbf {\bibinfo {volume} {16}},\ \bibinfo {pages}
		{448} (\bibinfo {year} {2020})}\BibitemShut {NoStop}%
	\bibitem [{\citenamefont {Tanaka}\ \emph {et~al.}(2019)\citenamefont {Tanaka},
		\citenamefont {Tong}, \citenamefont {Shi},\ and\ \citenamefont
		{Russo}}]{Hajime2019}%
	\BibitemOpen
	\bibfield  {author} {\bibinfo {author} {\bibfnamefont {H.}~\bibnamefont
			{Tanaka}}, \bibinfo {author} {\bibfnamefont {H.}~\bibnamefont {Tong}},
		\bibinfo {author} {\bibfnamefont {R.}~\bibnamefont {Shi}},\ and\ \bibinfo
		{author} {\bibfnamefont {J.}~\bibnamefont {Russo}},\ }\bibfield  {title}
	{\bibinfo {title} {Revealing key structural features hidden in liquids and
			glasses},\ }\href {https://doi.org/10.1038/s42254-019-0053-3} {\bibfield
		{journal} {\bibinfo  {journal} {Nat. Rev. Phys.}\ }\textbf {\bibinfo {volume}
			{1}},\ \bibinfo {pages} {333} (\bibinfo {year} {2019})}\BibitemShut {NoStop}%
	\bibitem [{\citenamefont {Ishino}\ \emph {et~al.}(2025)\citenamefont {Ishino},
		\citenamefont {Hu},\ and\ \citenamefont {Tanaka}}]{ishino2025a}%
	\BibitemOpen
	\bibfield  {author} {\bibinfo {author} {\bibfnamefont {S.}~\bibnamefont
			{Ishino}}, \bibinfo {author} {\bibfnamefont {Y.-C.}\ \bibnamefont {Hu}},\
		and\ \bibinfo {author} {\bibfnamefont {H.}~\bibnamefont {Tanaka}},\
	}\bibfield  {title} {\bibinfo {title} {Microscopic structural origin of slow
			dynamics in glass-forming liquids},\ }\href
	{https://doi.org/10.1038/s41563-024-02068-8} {\bibfield  {journal} {\bibinfo
			{journal} {Nat. Mater.}\ }\textbf {\bibinfo {volume} {24}},\ \bibinfo {pages}
		{268} (\bibinfo {year} {2025})}\BibitemShut {NoStop}%
	\bibitem [{\citenamefont {Gao}\ \emph {et~al.}(2025)\citenamefont {Gao},
		\citenamefont {Yu}, \citenamefont {Schrøder},\ and\ \citenamefont
		{Dyre}}]{gao2025}%
	\BibitemOpen
	\bibfield  {author} {\bibinfo {author} {\bibfnamefont {L.}~\bibnamefont
			{Gao}}, \bibinfo {author} {\bibfnamefont {H.-B.}\ \bibnamefont {Yu}},
		\bibinfo {author} {\bibfnamefont {T.~B.}\ \bibnamefont {Schrøder}},\ and\
		\bibinfo {author} {\bibfnamefont {J.~C.}\ \bibnamefont {Dyre}},\ }\bibfield
	{title} {\bibinfo {title} {Unified percolation scenario for the $\alpha$ and
			$\beta$ processes in simple glass formers},\ }\href
	{https://doi.org/10.1038/s41567-024-02762-z} {\bibfield  {journal} {\bibinfo
			{journal} {Nat. Phys.}\ }\textbf {\bibinfo {volume} {21}},\ \bibinfo {pages}
		{471} (\bibinfo {year} {2025})}\BibitemShut {NoStop}%
	\bibitem [{\citenamefont {Frey}\ \emph {et~al.}(2025)\citenamefont {Frey},
		\citenamefont {Neuber}, \citenamefont {Riegler}, \citenamefont {Cornet},
		\citenamefont {Chushkin}, \citenamefont {Zontone}, \citenamefont {Ruschel},
		\citenamefont {Adam}, \citenamefont {Nabahat}, \citenamefont {Yang},
		\citenamefont {Shen}, \citenamefont {Westermeier}, \citenamefont {Sprung},
		\citenamefont {Cangialosi}, \citenamefont {Di~Lisio}, \citenamefont
		{Gallino}, \citenamefont {Busch}, \citenamefont {Ruta},\ and\ \citenamefont
		{Pineda}}]{frey2025}%
	\BibitemOpen
	\bibfield  {author} {\bibinfo {author} {\bibfnamefont {M.}~\bibnamefont
			{Frey}}, \bibinfo {author} {\bibfnamefont {N.}~\bibnamefont {Neuber}},
		\bibinfo {author} {\bibfnamefont {S.~S.}\ \bibnamefont {Riegler}}, \bibinfo
		{author} {\bibfnamefont {A.}~\bibnamefont {Cornet}}, \bibinfo {author}
		{\bibfnamefont {Y.}~\bibnamefont {Chushkin}}, \bibinfo {author}
		{\bibfnamefont {F.}~\bibnamefont {Zontone}}, \bibinfo {author} {\bibfnamefont
			{L.~M.}\ \bibnamefont {Ruschel}}, \bibinfo {author} {\bibfnamefont
			{B.}~\bibnamefont {Adam}}, \bibinfo {author} {\bibfnamefont {M.}~\bibnamefont
			{Nabahat}}, \bibinfo {author} {\bibfnamefont {F.}~\bibnamefont {Yang}},
		\bibinfo {author} {\bibfnamefont {J.}~\bibnamefont {Shen}}, \bibinfo {author}
		{\bibfnamefont {F.}~\bibnamefont {Westermeier}}, \bibinfo {author}
		{\bibfnamefont {M.}~\bibnamefont {Sprung}}, \bibinfo {author} {\bibfnamefont
			{D.}~\bibnamefont {Cangialosi}}, \bibinfo {author} {\bibfnamefont
			{V.}~\bibnamefont {Di~Lisio}}, \bibinfo {author} {\bibfnamefont
			{I.}~\bibnamefont {Gallino}}, \bibinfo {author} {\bibfnamefont
			{R.}~\bibnamefont {Busch}}, \bibinfo {author} {\bibfnamefont
			{B.}~\bibnamefont {Ruta}},\ and\ \bibinfo {author} {\bibfnamefont
			{E.}~\bibnamefont {Pineda}},\ }\bibfield  {title} {\bibinfo {title}
		{Liquid-like versus stress-driven dynamics in a metallic glass former
			observed by temperature scanning {{X-ray}} photon correlation spectroscopy},\
	}\href {https://doi.org/10.1038/s41467-025-59767-2} {\bibfield  {journal}
		{\bibinfo  {journal} {Nat. Commun.}\ }\textbf {\bibinfo {volume} {16}},\
		\bibinfo {pages} {4429} (\bibinfo {year} {2025})}\BibitemShut {NoStop}%
	\bibitem [{\citenamefont {Berthier}\ and\ \citenamefont
		{Biroli}(2011)}]{Ludovic2011}%
	\BibitemOpen
	\bibfield  {author} {\bibinfo {author} {\bibfnamefont {L.}~\bibnamefont
			{Berthier}}\ and\ \bibinfo {author} {\bibfnamefont {G.}~\bibnamefont
			{Biroli}},\ }\bibfield  {title} {\bibinfo {title} {Theoretical perspective on
			the glass transition and amorphous materials},\ }\href
	{https://doi.org/10.1103/RevModPhys.83.587} {\bibfield  {journal} {\bibinfo
			{journal} {Rev. Mod. Phys.}\ }\textbf {\bibinfo {volume} {83}},\ \bibinfo
		{pages} {587} (\bibinfo {year} {2011})}\BibitemShut {NoStop}%
	\bibitem [{\citenamefont {Dyre}(2024)}]{dyre2024}%
	\BibitemOpen
	\bibfield  {author} {\bibinfo {author} {\bibfnamefont {J.~C.}\ \bibnamefont
			{Dyre}},\ }\bibfield  {title} {\bibinfo {title} {Solid-that-flows: Picture of
			glass-forming liquids},\ }\href {https://doi.org/10.1021/acs.jpclett.3c03308}
	{\bibfield  {journal} {\bibinfo  {journal} {J. Phys. Chem. Lett.}\ }\textbf
		{\bibinfo {volume} {15}},\ \bibinfo {pages} {1603} (\bibinfo {year}
		{2024})}\BibitemShut {NoStop}%
	\bibitem [{\citenamefont {Lunkenheimer}\ \emph {et~al.}(2023)\citenamefont
		{Lunkenheimer}, \citenamefont {Loidl}, \citenamefont {Riechers},
		\citenamefont {Zaccone},\ and\ \citenamefont {Samwer}}]{lunkenheimer2023}%
	\BibitemOpen
	\bibfield  {author} {\bibinfo {author} {\bibfnamefont {P.}~\bibnamefont
			{Lunkenheimer}}, \bibinfo {author} {\bibfnamefont {A.}~\bibnamefont {Loidl}},
		\bibinfo {author} {\bibfnamefont {B.}~\bibnamefont {Riechers}}, \bibinfo
		{author} {\bibfnamefont {A.}~\bibnamefont {Zaccone}},\ and\ \bibinfo {author}
		{\bibfnamefont {K.}~\bibnamefont {Samwer}},\ }\bibfield  {title} {\bibinfo
		{title} {Thermal expansion and the glass transition},\ }\href
	{https://doi.org/10.1038/s41567-022-01920-5} {\bibfield  {journal} {\bibinfo
			{journal} {Nat. Phys.}\ }\textbf {\bibinfo {volume} {19}},\ \bibinfo {pages}
		{694} (\bibinfo {year} {2023})}\BibitemShut {NoStop}%
	\bibitem [{\citenamefont {Granata}\ \emph {et~al.}(2014)\citenamefont
		{Granata}, \citenamefont {Fischer}, \citenamefont {Wessels},\ and\
		\citenamefont {L{\"o}ffler}}]{granata2014}%
	\BibitemOpen
	\bibfield  {author} {\bibinfo {author} {\bibfnamefont {D.}~\bibnamefont
			{Granata}}, \bibinfo {author} {\bibfnamefont {E.}~\bibnamefont {Fischer}},
		\bibinfo {author} {\bibfnamefont {V.}~\bibnamefont {Wessels}},\ and\ \bibinfo
		{author} {\bibfnamefont {J.}~\bibnamefont {L{\"o}ffler}},\ }\bibfield
	{title} {\bibinfo {title} {Fluxing of {{Pd}}-{{Si}}-{{Cu}} bulk metallic
			glass and the role of cooling rate and purification},\ }\href
	{https://doi.org/10.1016/j.actamat.2014.03.008} {\bibfield  {journal}
		{\bibinfo  {journal} {Acta Mater.}\ }\textbf {\bibinfo {volume} {71}},\
		\bibinfo {pages} {145} (\bibinfo {year} {2014})}\BibitemShut {NoStop}%
	\bibitem [{\citenamefont {Atila}\ \emph {et~al.}(2025)\citenamefont {Atila},
		\citenamefont {Sukhomlinov}, \citenamefont {Honecker},\ and\ \citenamefont
		{M{\"u}ser}}]{atila2025}%
	\BibitemOpen
	\bibfield  {author} {\bibinfo {author} {\bibfnamefont {A.}~\bibnamefont
			{Atila}}, \bibinfo {author} {\bibfnamefont {S.~V.}\ \bibnamefont
			{Sukhomlinov}}, \bibinfo {author} {\bibfnamefont {M.~J.}\ \bibnamefont
			{Honecker}},\ and\ \bibinfo {author} {\bibfnamefont {M.~H.}\ \bibnamefont
			{M{\"u}ser}},\ }\bibfield  {title} {\bibinfo {title} {Plasticity of metallic
			glasses dictated by their state at the fragile-to-strong transition
			temperature},\ }\href {https://doi.org/10.1016/j.actamat.2025.120753}
	{\bibfield  {journal} {\bibinfo  {journal} {Acta Mater.}\ }\textbf {\bibinfo
			{volume} {286}},\ \bibinfo {pages} {120753} (\bibinfo {year}
		{2025})}\BibitemShut {NoStop}%
	\bibitem [{\citenamefont {Schawe}\ and\ \citenamefont
		{L{\"o}ffler}(2022)}]{schawe2022}%
	\BibitemOpen
	\bibfield  {author} {\bibinfo {author} {\bibfnamefont {J.~E.}\ \bibnamefont
			{Schawe}}\ and\ \bibinfo {author} {\bibfnamefont {J.~F.}\ \bibnamefont
			{L{\"o}ffler}},\ }\bibfield  {title} {\bibinfo {title} {Kinetics of structure
			formation in the vicinity of the glass transition},\ }\href
	{https://doi.org/10.1016/j.actamat.2022.117630} {\bibfield  {journal}
		{\bibinfo  {journal} {Acta Materialia}\ }\textbf {\bibinfo {volume} {226}},\
		\bibinfo {pages} {117630} (\bibinfo {year} {2022})}\BibitemShut {NoStop}%
	\bibitem [{\citenamefont {Da~Silva~Pinto}\ \emph {et~al.}(2024)\citenamefont
		{Da~Silva~Pinto}, \citenamefont {Daum}, \citenamefont {R{\"o}sner},\ and\
		\citenamefont {Wilde}}]{da2024}%
	\BibitemOpen
	\bibfield  {author} {\bibinfo {author} {\bibfnamefont {M.~W.}\ \bibnamefont
			{Da~Silva~Pinto}}, \bibinfo {author} {\bibfnamefont {L.}~\bibnamefont
			{Daum}}, \bibinfo {author} {\bibfnamefont {H.}~\bibnamefont {R{\"o}sner}},\
		and\ \bibinfo {author} {\bibfnamefont {G.}~\bibnamefont {Wilde}},\ }\bibfield
	{title} {\bibinfo {title} {Correlations between shadow glass transition,
			enthalpy recovery and medium range order in a {{Pd$_{40}$Ni$_{40}$P$_{20}$}}
			bulk metallic glass},\ }\href {https://doi.org/10.1016/j.actamat.2024.120034}
	{\bibfield  {journal} {\bibinfo  {journal} {Acta Mater.}\ }\textbf {\bibinfo
			{volume} {275}},\ \bibinfo {pages} {120034} (\bibinfo {year}
		{2024})}\BibitemShut {NoStop}%
	\bibitem [{\citenamefont {Spieckermann}\ \emph {et~al.}(2022)\citenamefont
		{Spieckermann}, \citenamefont {{\c S}opu}, \citenamefont {Soprunyuk},
		\citenamefont {Kerber}, \citenamefont {Bednar{\v c}{\'i}k}, \citenamefont
		{Sch{\"o}kel}, \citenamefont {Rezvan}, \citenamefont {Ketov}, \citenamefont
		{Sarac}, \citenamefont {Schafler},\ and\ \citenamefont {Eckert}}]{spie2022}%
	\BibitemOpen
	\bibfield  {author} {\bibinfo {author} {\bibfnamefont {F.}~\bibnamefont
			{Spieckermann}}, \bibinfo {author} {\bibfnamefont {D.}~\bibnamefont {{\c
					S}opu}}, \bibinfo {author} {\bibfnamefont {V.}~\bibnamefont {Soprunyuk}},
		\bibinfo {author} {\bibfnamefont {M.~B.}\ \bibnamefont {Kerber}}, \bibinfo
		{author} {\bibfnamefont {J.}~\bibnamefont {Bednar{\v c}{\'i}k}}, \bibinfo
		{author} {\bibfnamefont {A.}~\bibnamefont {Sch{\"o}kel}}, \bibinfo {author}
		{\bibfnamefont {A.}~\bibnamefont {Rezvan}}, \bibinfo {author} {\bibfnamefont
			{S.}~\bibnamefont {Ketov}}, \bibinfo {author} {\bibfnamefont
			{B.}~\bibnamefont {Sarac}}, \bibinfo {author} {\bibfnamefont
			{E.}~\bibnamefont {Schafler}},\ and\ \bibinfo {author} {\bibfnamefont
			{J.}~\bibnamefont {Eckert}},\ }\bibfield  {title} {\bibinfo {title}
		{Structure-dynamics relationships in cryogenically deformed bulk metallic
			glass},\ }\href {https://doi.org/10.1038/s41467-021-27661-2} {\bibfield
		{journal} {\bibinfo  {journal} {Nat. Commun.}\ }\textbf {\bibinfo {volume}
			{13}},\ \bibinfo {pages} {127} (\bibinfo {year} {2022})}\BibitemShut
	{NoStop}%
	\bibitem [{\citenamefont {Böhmer}\ \emph {et~al.}(2024)\citenamefont
		{Böhmer}, \citenamefont {Gabriel}, \citenamefont {Costigliola},
		\citenamefont {Kociok}, \citenamefont {Hecksher}, \citenamefont {Dyre},\ and\
		\citenamefont {Blochowicz}}]{bohmer2024}%
	\BibitemOpen
	\bibfield  {author} {\bibinfo {author} {\bibfnamefont {T.}~\bibnamefont
			{Böhmer}}, \bibinfo {author} {\bibfnamefont {J.~P.}\ \bibnamefont
			{Gabriel}}, \bibinfo {author} {\bibfnamefont {L.}~\bibnamefont
			{Costigliola}}, \bibinfo {author} {\bibfnamefont {J.-N.}\ \bibnamefont
			{Kociok}}, \bibinfo {author} {\bibfnamefont {T.}~\bibnamefont {Hecksher}},
		\bibinfo {author} {\bibfnamefont {J.~C.}\ \bibnamefont {Dyre}},\ and\
		\bibinfo {author} {\bibfnamefont {T.}~\bibnamefont {Blochowicz}},\ }\bibfield
	{title} {\bibinfo {title} {Time reversibility during the ageing of
			materials},\ }\href {https://doi.org/10.1038/s41567-023-02366-z} {\bibfield
		{journal} {\bibinfo  {journal} {Nat. Phys.}\ }\textbf {\bibinfo {volume}
			{20}},\ \bibinfo {pages} {637} (\bibinfo {year} {2024})}\BibitemShut
	{NoStop}%
	\bibitem [{\citenamefont {Saini}\ \emph {et~al.}(2024)\citenamefont {Saini},
		\citenamefont {Zhao}, \citenamefont {Li}, \citenamefont {Zhang},
		\citenamefont {Ramamurty},\ and\ \citenamefont {Narayan}}]{saini2024}%
	\BibitemOpen
	\bibfield  {author} {\bibinfo {author} {\bibfnamefont {P.}~\bibnamefont
			{Saini}}, \bibinfo {author} {\bibfnamefont {Y.}~\bibnamefont {Zhao}},
		\bibinfo {author} {\bibfnamefont {B.}~\bibnamefont {Li}}, \bibinfo {author}
		{\bibfnamefont {L.}~\bibnamefont {Zhang}}, \bibinfo {author} {\bibfnamefont
			{U.}~\bibnamefont {Ramamurty}},\ and\ \bibinfo {author} {\bibfnamefont
			{R.}~\bibnamefont {Narayan}},\ }\bibfield  {title} {\bibinfo {title}
		{Temperature dependence of pressure sensitive flow in bulk metallic glass
			composites},\ }\href {https://doi.org/10.1016/j.jmst.2023.09.028} {\bibfield
		{journal} {\bibinfo  {journal} {J. Mater. Sci. Technol.}\ }\textbf {\bibinfo
			{volume} {181}},\ \bibinfo {pages} {165} (\bibinfo {year}
		{2024})}\BibitemShut {NoStop}%
	\bibitem [{\citenamefont {L\"uttich}\ \emph {et~al.}(2018)\citenamefont
		{L\"uttich}, \citenamefont {Giordano}, \citenamefont {Le~Floch},
		\citenamefont {Pineda}, \citenamefont {Zontone}, \citenamefont {Luo},
		\citenamefont {Samwer},\ and\ \citenamefont {Ruta}}]{Martin2018}%
	\BibitemOpen
	\bibfield  {author} {\bibinfo {author} {\bibfnamefont {M.}~\bibnamefont
			{L\"uttich}}, \bibinfo {author} {\bibfnamefont {V.~M.}\ \bibnamefont
			{Giordano}}, \bibinfo {author} {\bibfnamefont {S.}~\bibnamefont {Le~Floch}},
		\bibinfo {author} {\bibfnamefont {E.}~\bibnamefont {Pineda}}, \bibinfo
		{author} {\bibfnamefont {F.}~\bibnamefont {Zontone}}, \bibinfo {author}
		{\bibfnamefont {Y.}~\bibnamefont {Luo}}, \bibinfo {author} {\bibfnamefont
			{K.}~\bibnamefont {Samwer}},\ and\ \bibinfo {author} {\bibfnamefont
			{B.}~\bibnamefont {Ruta}},\ }\bibfield  {title} {\bibinfo {title} {Anti-aging
			in ultrastable metallic glasses},\ }\href
	{https://doi.org/10.1103/PhysRevLett.120.135504} {\bibfield  {journal}
		{\bibinfo  {journal} {Phys. Rev. Lett.}\ }\textbf {\bibinfo {volume} {120}},\
		\bibinfo {pages} {135504} (\bibinfo {year} {2018})}\BibitemShut {NoStop}%
	\bibitem [{\citenamefont {Wang}\ \emph {et~al.}(2018)\citenamefont {Wang},
		\citenamefont {Xu}, \citenamefont {Wang},\ and\ \citenamefont
		{Guan}}]{Guan2018}%
	\BibitemOpen
	\bibfield  {author} {\bibinfo {author} {\bibfnamefont {L.}~\bibnamefont
			{Wang}}, \bibinfo {author} {\bibfnamefont {N.}~\bibnamefont {Xu}}, \bibinfo
		{author} {\bibfnamefont {W.~H.}\ \bibnamefont {Wang}},\ and\ \bibinfo
		{author} {\bibfnamefont {P.}~\bibnamefont {Guan}},\ }\bibfield  {title}
	{\bibinfo {title} {Revealing the link between structural relaxation and
			dynamic heterogeneity in glass-forming liquids},\ }\href
	{https://doi.org/10.1103/PhysRevLett.120.125502} {\bibfield  {journal}
		{\bibinfo  {journal} {Phys. Rev. Lett.}\ }\textbf {\bibinfo {volume} {120}},\
		\bibinfo {pages} {125502} (\bibinfo {year} {2018})}\BibitemShut {NoStop}%
	\bibitem [{\citenamefont {Peter~G.}(2009)}]{peter2009}%
	\BibitemOpen
	\bibfield  {author} {\bibinfo {author} {\bibfnamefont {W.}~\bibnamefont
			{Peter~G.}},\ }\bibfield  {title} {\bibinfo {title} {Spatiotemporal
			structures in aging and rejuvenating glasses},\ }\href
	{https://doi.org/10.1073/pnas.0812418106} {\bibfield  {journal} {\bibinfo
			{journal} {Proc. Nat. Acad. Sci.}\ }\textbf {\bibinfo {volume} {106}},\
		\bibinfo {pages} {1353} (\bibinfo {year} {2009})}\BibitemShut {NoStop}%
	\bibitem [{\citenamefont {Schawe}\ and\ \citenamefont
		{L{\"o}ffler}(2019)}]{schawe2019a}%
	\BibitemOpen
	\bibfield  {author} {\bibinfo {author} {\bibfnamefont {J.~E.~K.}\
			\bibnamefont {Schawe}}\ and\ \bibinfo {author} {\bibfnamefont {J.~F.}\
			\bibnamefont {L{\"o}ffler}},\ }\bibfield  {title} {\bibinfo {title}
		{Existence of multiple critical cooling rates which generate different types
			of monolithic metallic glass},\ }\href
	{https://doi.org/10.1038/s41467-018-07930-3} {\bibfield  {journal} {\bibinfo
			{journal} {Nat. Commun.}\ }\textbf {\bibinfo {volume} {10}},\ \bibinfo
		{pages} {1337} (\bibinfo {year} {2019})}\BibitemShut {NoStop}%
	\bibitem [{\citenamefont {Greer}(2015)}]{greer2015}%
	\BibitemOpen
	\bibfield  {author} {\bibinfo {author} {\bibfnamefont {A.~L.}\ \bibnamefont
			{Greer}},\ }\bibfield  {title} {\bibinfo {title} {New horizons for glass
			formation and stability},\ }\href {https://doi.org/10.1038/nmat4292}
	{\bibfield  {journal} {\bibinfo  {journal} {Nat. Mater.}\ }\textbf {\bibinfo
			{volume} {14}},\ \bibinfo {pages} {542} (\bibinfo {year} {2015})}\BibitemShut
	{NoStop}%
	\bibitem [{\citenamefont {Cheng}\ \emph {et~al.}(2022)\citenamefont {Cheng},
		\citenamefont {Yang}, \citenamefont {Wang}, \citenamefont {Dimitriadis},
		\citenamefont {Schumacher}, \citenamefont {Zhang}, \citenamefont
		{M{\"u}ller}, \citenamefont {Amini}, \citenamefont {Yang}, \citenamefont
		{Schoekel}, \citenamefont {Pries}, \citenamefont {Mazzarello}, \citenamefont
		{Wuttig}, \citenamefont {Yu},\ and\ \citenamefont {Wei}}]{chengnc2022}%
	\BibitemOpen
	\bibfield  {author} {\bibinfo {author} {\bibfnamefont {Y.~D.}\ \bibnamefont
			{Cheng}}, \bibinfo {author} {\bibfnamefont {Q.}~\bibnamefont {Yang}},
		\bibinfo {author} {\bibfnamefont {J.~J.}\ \bibnamefont {Wang}}, \bibinfo
		{author} {\bibfnamefont {T.}~\bibnamefont {Dimitriadis}}, \bibinfo {author}
		{\bibfnamefont {M.}~\bibnamefont {Schumacher}}, \bibinfo {author}
		{\bibfnamefont {H.~R.}\ \bibnamefont {Zhang}}, \bibinfo {author}
		{\bibfnamefont {M.~J.}\ \bibnamefont {M{\"u}ller}}, \bibinfo {author}
		{\bibfnamefont {N.}~\bibnamefont {Amini}}, \bibinfo {author} {\bibfnamefont
			{F.}~\bibnamefont {Yang}}, \bibinfo {author} {\bibfnamefont {A.}~\bibnamefont
			{Schoekel}}, \bibinfo {author} {\bibfnamefont {J.}~\bibnamefont {Pries}},
		\bibinfo {author} {\bibfnamefont {R.}~\bibnamefont {Mazzarello}}, \bibinfo
		{author} {\bibfnamefont {M.}~\bibnamefont {Wuttig}}, \bibinfo {author}
		{\bibfnamefont {H.-B.}\ \bibnamefont {Yu}},\ and\ \bibinfo {author}
		{\bibfnamefont {S.}~\bibnamefont {Wei}},\ }\bibfield  {title} {\bibinfo
		{title} {Highly tunable {$\beta$}-relaxation enables the tailoring of
			crystallization in phase-change materials},\ }\href
	{https://doi.org/10.1038/s41467-022-35005-x} {\bibfield  {journal} {\bibinfo
			{journal} {Nat. Commun.}\ }\textbf {\bibinfo {volume} {13}},\ \bibinfo
		{pages} {7352} (\bibinfo {year} {2022})}\BibitemShut {NoStop}%
	\bibitem [{\citenamefont {{L{\'a}zaro-L{\'a}zaro}}\ \emph
		{et~al.}(2019)\citenamefont {{L{\'a}zaro-L{\'a}zaro}}, \citenamefont
		{{Perera-Burgos}}, \citenamefont {Laermann}, \citenamefont {Sentjabrskaja},
		\citenamefont {{P{\'e}rez-{\'A}ngel}}, \citenamefont {Laurati}, \citenamefont
		{Egelhaaf}, \citenamefont {{Medina-Noyola}}, \citenamefont {Voigtmann},
		\citenamefont {{Casta{\~n}eda-Priego}},\ and\ \citenamefont
		{{Elizondo-Aguilera}}}]{lazaro2019}%
	\BibitemOpen
	\bibfield  {author} {\bibinfo {author} {\bibfnamefont {E.}~\bibnamefont
			{{L{\'a}zaro-L{\'a}zaro}}}, \bibinfo {author} {\bibfnamefont {J.~A.}\
			\bibnamefont {{Perera-Burgos}}}, \bibinfo {author} {\bibfnamefont
			{P.}~\bibnamefont {Laermann}}, \bibinfo {author} {\bibfnamefont
			{T.}~\bibnamefont {Sentjabrskaja}}, \bibinfo {author} {\bibfnamefont
			{G.}~\bibnamefont {{P{\'e}rez-{\'A}ngel}}}, \bibinfo {author} {\bibfnamefont
			{M.}~\bibnamefont {Laurati}}, \bibinfo {author} {\bibfnamefont {S.~U.}\
			\bibnamefont {Egelhaaf}}, \bibinfo {author} {\bibfnamefont {M.}~\bibnamefont
			{{Medina-Noyola}}}, \bibinfo {author} {\bibfnamefont {T.}~\bibnamefont
			{Voigtmann}}, \bibinfo {author} {\bibfnamefont {R.}~\bibnamefont
			{{Casta{\~n}eda-Priego}}},\ and\ \bibinfo {author} {\bibfnamefont {L.~F.}\
			\bibnamefont {{Elizondo-Aguilera}}},\ }\bibfield  {title} {\bibinfo {title}
		{Glassy dynamics in asymmetric binary mixtures of hard spheres},\ }\href
	{https://doi.org/10.1103/PhysRevE.99.042603} {\bibfield  {journal} {\bibinfo
			{journal} {Phys. Rev. E}\ }\textbf {\bibinfo {volume} {99}},\ \bibinfo
		{pages} {42603} (\bibinfo {year} {2019})}\BibitemShut {NoStop}%
	\bibitem [{\citenamefont {Yu}\ \emph {et~al.}(2024)\citenamefont {Yu},
		\citenamefont {Gao}, \citenamefont {Gao},\ and\ \citenamefont
		{Samwer}}]{yunsr2024}%
	\BibitemOpen
	\bibfield  {author} {\bibinfo {author} {\bibfnamefont {H.-B.}\ \bibnamefont
			{Yu}}, \bibinfo {author} {\bibfnamefont {L.}~\bibnamefont {Gao}}, \bibinfo
		{author} {\bibfnamefont {J.-Q.}\ \bibnamefont {Gao}},\ and\ \bibinfo {author}
		{\bibfnamefont {K.}~\bibnamefont {Samwer}},\ }\bibfield  {title} {\bibinfo
		{title} {Universal origin of glassy relaxation as recognized by configuration
			pattern matching},\ }\href {https://doi.org/10.1093/nsr/nwae091} {\bibfield
		{journal} {\bibinfo  {journal} {Natl. Sci. Rev.}\ }\textbf {\bibinfo {volume}
			{11}},\ \bibinfo {pages} {nwae091} (\bibinfo {year} {2024})}\BibitemShut
	{NoStop}%
	\bibitem [{\citenamefont {Zhou}\ \emph {et~al.}(2023)\citenamefont {Zhou},
		\citenamefont {Sun}, \citenamefont {Gao}, \citenamefont {Wang},\ and\
		\citenamefont {Yu}}]{zhouacta2023}%
	\BibitemOpen
	\bibfield  {author} {\bibinfo {author} {\bibfnamefont {Z.-Y.}\ \bibnamefont
			{Zhou}}, \bibinfo {author} {\bibfnamefont {Y.}~\bibnamefont {Sun}}, \bibinfo
		{author} {\bibfnamefont {L.}~\bibnamefont {Gao}}, \bibinfo {author}
		{\bibfnamefont {Y.-J.}\ \bibnamefont {Wang}},\ and\ \bibinfo {author}
		{\bibfnamefont {H.-B.}\ \bibnamefont {Yu}},\ }\bibfield  {title} {\bibinfo
		{title} {Fundamental links between shear transformation, {$\beta$}
			relaxation, and string-like motion in metallic glasses},\ }\href
	{https://doi.org/10.1016/j.actamat.2023.118701} {\bibfield  {journal}
		{\bibinfo  {journal} {Acta Mater.}\ }\textbf {\bibinfo {volume} {246}},\
		\bibinfo {pages} {118701} (\bibinfo {year} {2023})}\BibitemShut {NoStop}%
	\bibitem [{\citenamefont {Parisi}(1983)}]{parisi1983}%
	\BibitemOpen
	\bibfield  {author} {\bibinfo {author} {\bibfnamefont {G.}~\bibnamefont
			{Parisi}},\ }\bibfield  {title} {\bibinfo {title} {Order parameter for
			spin-glasses},\ }\href {https://doi.org/10.1103/PhysRevLett.50.1946}
	{\bibfield  {journal} {\bibinfo  {journal} {Phys. Rev. Lett.}\ }\textbf
		{\bibinfo {volume} {50}},\ \bibinfo {pages} {1946} (\bibinfo {year}
		{1983})}\BibitemShut {NoStop}%
	\bibitem [{\citenamefont {Guiselin}\ \emph {et~al.}(2020)\citenamefont
		{Guiselin}, \citenamefont {Tarjus},\ and\ \citenamefont
		{Berthier}}]{gui2020}%
	\BibitemOpen
	\bibfield  {author} {\bibinfo {author} {\bibfnamefont {B.}~\bibnamefont
			{Guiselin}}, \bibinfo {author} {\bibfnamefont {G.}~\bibnamefont {Tarjus}},\
		and\ \bibinfo {author} {\bibfnamefont {L.}~\bibnamefont {Berthier}},\
	}\bibfield  {title} {\bibinfo {title} {On the overlap between configurations
			in glassy liquids},\ }\href {https://doi.org/10.1063/5.0022614} {\bibfield
		{journal} {\bibinfo  {journal} {J. Chem. Phys.}\ }\textbf {\bibinfo {volume}
			{153}},\ \bibinfo {pages} {224502} (\bibinfo {year} {2020})}\BibitemShut
	{NoStop}%
	\bibitem [{\citenamefont {Crouse}(2016)}]{crouse2016b}%
	\BibitemOpen
	\bibfield  {author} {\bibinfo {author} {\bibfnamefont {D.~F.}\ \bibnamefont
			{Crouse}},\ }\bibfield  {title} {\bibinfo {title} {On implementing {{2D}}
			rectangular assignment algorithms},\ }\href
	{https://doi.org/10.1109/taes.2016.140952} {\bibfield  {journal} {\bibinfo
			{journal} {IEEE Trans. Aerosp. Electron. Syst.}\ }\textbf {\bibinfo {volume}
			{52}},\ \bibinfo {pages} {1679} (\bibinfo {year} {2016})}\BibitemShut
	{NoStop}%
	\bibitem [{\citenamefont {Tanaka}(2003)}]{Tanaka2003}%
	\BibitemOpen
	\bibfield  {author} {\bibinfo {author} {\bibfnamefont {H.}~\bibnamefont
			{Tanaka}},\ }\bibfield  {title} {\bibinfo {title} {Relation between
			thermodynamics and kinetics of glass-forming liquids},\ }\href
	{https://doi.org/10.1103/PhysRevLett.90.055701} {\bibfield  {journal}
		{\bibinfo  {journal} {Phys. Rev. Lett.}\ }\textbf {\bibinfo {volume} {90}},\
		\bibinfo {pages} {055701} (\bibinfo {year} {2003})}\BibitemShut {NoStop}%
	\bibitem [{\citenamefont {Hecksher}\ \emph {et~al.}(2008)\citenamefont
		{Hecksher}, \citenamefont {Nielsen}, \citenamefont {Olsen},\ and\
		\citenamefont {Dyre}}]{dyre2008}%
	\BibitemOpen
	\bibfield  {author} {\bibinfo {author} {\bibfnamefont {T.}~\bibnamefont
			{Hecksher}}, \bibinfo {author} {\bibfnamefont {A.~I.}\ \bibnamefont
			{Nielsen}}, \bibinfo {author} {\bibfnamefont {N.~B.}\ \bibnamefont {Olsen}},\
		and\ \bibinfo {author} {\bibfnamefont {J.~C.}\ \bibnamefont {Dyre}},\
	}\bibfield  {title} {\bibinfo {title} {Little evidence for dynamic
			divergences in ultraviscous molecular liquids},\ }\href
	{https://doi.org/10.1038/nphys1033} {\bibfield  {journal} {\bibinfo
			{journal} {Nat. Phys.}\ }\textbf {\bibinfo {volume} {4}},\ \bibinfo {pages}
		{737} (\bibinfo {year} {2008})}\BibitemShut {NoStop}%
	\bibitem [{\citenamefont {Adhikari}\ \emph {et~al.}(2023)\citenamefont
		{Adhikari}, \citenamefont {Karmakar},\ and\ \citenamefont
		{Sastry}}]{Adhikari2023}%
	\BibitemOpen
	\bibfield  {author} {\bibinfo {author} {\bibfnamefont {M.}~\bibnamefont
			{Adhikari}}, \bibinfo {author} {\bibfnamefont {S.}~\bibnamefont {Karmakar}},\
		and\ \bibinfo {author} {\bibfnamefont {S.}~\bibnamefont {Sastry}},\
	}\bibfield  {title} {\bibinfo {title} {Dependence of the glass transition and
			jamming densities on spatial dimension},\ }\href
	{https://doi.org/10.1103/PhysRevLett.131.168202} {\bibfield  {journal}
		{\bibinfo  {journal} {Phys. Rev. Lett.}\ }\textbf {\bibinfo {volume} {131}},\
		\bibinfo {pages} {168202} (\bibinfo {year} {2023})}\BibitemShut {NoStop}%
	\bibitem [{\citenamefont {Yoon}\ and\ \citenamefont
		{McKenna}(2018)}]{yoon2018}%
	\BibitemOpen
	\bibfield  {author} {\bibinfo {author} {\bibfnamefont {H.}~\bibnamefont
			{Yoon}}\ and\ \bibinfo {author} {\bibfnamefont {G.~B.}\ \bibnamefont
			{McKenna}},\ }\bibfield  {title} {\bibinfo {title} {Testing the paradigm of
			an ideal glass transition: {{Dynamics}} of an ultrastable polymeric glass},\
	}\href {https://doi.org/10.1126/sciadv.aau5423} {\bibfield  {journal}
		{\bibinfo  {journal} {Sci. Adv.}\ }\textbf {\bibinfo {volume} {4}},\ \bibinfo
		{pages} {eaau5423} (\bibinfo {year} {2018})}\BibitemShut {NoStop}%
	\bibitem [{\citenamefont {Monnier}\ \emph {et~al.}(2021)\citenamefont
		{Monnier}, \citenamefont {Colmenero}, \citenamefont {Wolf},\ and\
		\citenamefont {Cangialosi}}]{Monnier2021}%
	\BibitemOpen
	\bibfield  {author} {\bibinfo {author} {\bibfnamefont {X.}~\bibnamefont
			{Monnier}}, \bibinfo {author} {\bibfnamefont {J.}~\bibnamefont {Colmenero}},
		\bibinfo {author} {\bibfnamefont {M.}~\bibnamefont {Wolf}},\ and\ \bibinfo
		{author} {\bibfnamefont {D.}~\bibnamefont {Cangialosi}},\ }\bibfield  {title}
	{\bibinfo {title} {Reaching the ideal glass in polymer spheres:
			Thermodynamics and vibrational density of states},\ }\href
	{https://doi.org/10.1103/PhysRevLett.126.118004} {\bibfield  {journal}
		{\bibinfo  {journal} {Phys. Rev. Lett.}\ }\textbf {\bibinfo {volume} {126}},\
		\bibinfo {pages} {118004} (\bibinfo {year} {2021})}\BibitemShut {NoStop}%
	\bibitem [{\citenamefont {Cammarota}\ and\ \citenamefont
		{Biroli}(2012)}]{cammarota2012}%
	\BibitemOpen
	\bibfield  {author} {\bibinfo {author} {\bibfnamefont {C.}~\bibnamefont
			{Cammarota}}\ and\ \bibinfo {author} {\bibfnamefont {G.}~\bibnamefont
			{Biroli}},\ }\bibfield  {title} {\bibinfo {title} {Ideal glass transitions by
			random pinning},\ }\href {https://doi.org/10.1073/pnas.1111582109} {\bibfield
		{journal} {\bibinfo  {journal} {Proc. Nat. Acad. Sci.}\ }\textbf {\bibinfo
			{volume} {109}},\ \bibinfo {pages} {8850} (\bibinfo {year}
		{2012})}\BibitemShut {NoStop}%
	\bibitem [{\citenamefont {Ozawa}\ \emph {et~al.}(2018)\citenamefont {Ozawa},
		\citenamefont {Ikeda}, \citenamefont {Miyazaki},\ and\ \citenamefont
		{Kob}}]{ozawa2018a}%
	\BibitemOpen
	\bibfield  {author} {\bibinfo {author} {\bibfnamefont {M.}~\bibnamefont
			{Ozawa}}, \bibinfo {author} {\bibfnamefont {A.}~\bibnamefont {Ikeda}},
		\bibinfo {author} {\bibfnamefont {K.}~\bibnamefont {Miyazaki}},\ and\
		\bibinfo {author} {\bibfnamefont {W.}~\bibnamefont {Kob}},\ }\bibfield
	{title} {\bibinfo {title} {Ideal glass states are not purely vibrational:
			{{Insight}} from randomly pinned glasses},\ }\href
	{https://doi.org/10.1103/PhysRevLett.121.205501} {\bibfield  {journal}
		{\bibinfo  {journal} {Phys. Rev. Lett.}\ }\textbf {\bibinfo {volume} {121}},\
		\bibinfo {pages} {205501} (\bibinfo {year} {2018})}\BibitemShut {NoStop}%
	\bibitem [{\citenamefont {Kob}\ and\ \citenamefont {Andersen}(1993)}]{Kob1993}%
	\BibitemOpen
	\bibfield  {author} {\bibinfo {author} {\bibfnamefont {W.}~\bibnamefont
			{Kob}}\ and\ \bibinfo {author} {\bibfnamefont {H.~C.}\ \bibnamefont
			{Andersen}},\ }\bibfield  {title} {\bibinfo {title} {Kinetic lattice-gas
			model of cage effects in high-density liquids and a test of mode-coupling
			theory of the ideal-glass transition},\ }\href
	{https://doi.org/10.1103/PhysRevE.48.4364} {\bibfield  {journal} {\bibinfo
			{journal} {Phys. Rev. E}\ }\textbf {\bibinfo {volume} {48}},\ \bibinfo
		{pages} {4364} (\bibinfo {year} {1993})}\BibitemShut {NoStop}%
	\bibitem [{\citenamefont {Stein}\ and\ \citenamefont
		{Andersen}(2008)}]{Andersen2008}%
	\BibitemOpen
	\bibfield  {author} {\bibinfo {author} {\bibfnamefont {R.~S.~L.}\
			\bibnamefont {Stein}}\ and\ \bibinfo {author} {\bibfnamefont {H.~C.}\
			\bibnamefont {Andersen}},\ }\bibfield  {title} {\bibinfo {title} {Scaling
			analysis of dynamic heterogeneity in a supercooled lennard-jones liquid},\
	}\href {https://doi.org/10.1103/PhysRevLett.101.267802} {\bibfield  {journal}
		{\bibinfo  {journal} {Phys. Rev. Lett.}\ }\textbf {\bibinfo {volume} {101}},\
		\bibinfo {pages} {267802} (\bibinfo {year} {2008})}\BibitemShut {NoStop}%
	\bibitem [{\citenamefont {Thompson}\ \emph {et~al.}(2022)\citenamefont
		{Thompson}, \citenamefont {Aktulga}, \citenamefont {Berger}, \citenamefont
		{Bolintineanu}, \citenamefont {Brown}, \citenamefont {Crozier}, \citenamefont
		{{in 't Veld}}, \citenamefont {Kohlmeyer}, \citenamefont {Moore},
		\citenamefont {Nguyen}, \citenamefont {Shan}, \citenamefont {Stevens},
		\citenamefont {Tranchida}, \citenamefont {Trott},\ and\ \citenamefont
		{Plimpton}}]{lmp22}%
	\BibitemOpen
	\bibfield  {author} {\bibinfo {author} {\bibfnamefont {A.~P.}\ \bibnamefont
			{Thompson}}, \bibinfo {author} {\bibfnamefont {H.~M.}\ \bibnamefont
			{Aktulga}}, \bibinfo {author} {\bibfnamefont {R.}~\bibnamefont {Berger}},
		\bibinfo {author} {\bibfnamefont {D.~S.}\ \bibnamefont {Bolintineanu}},
		\bibinfo {author} {\bibfnamefont {W.~M.}\ \bibnamefont {Brown}}, \bibinfo
		{author} {\bibfnamefont {P.~S.}\ \bibnamefont {Crozier}}, \bibinfo {author}
		{\bibfnamefont {P.~J.}\ \bibnamefont {{in 't Veld}}}, \bibinfo {author}
		{\bibfnamefont {A.}~\bibnamefont {Kohlmeyer}}, \bibinfo {author}
		{\bibfnamefont {S.~G.}\ \bibnamefont {Moore}}, \bibinfo {author}
		{\bibfnamefont {T.~D.}\ \bibnamefont {Nguyen}}, \bibinfo {author}
		{\bibfnamefont {R.}~\bibnamefont {Shan}}, \bibinfo {author} {\bibfnamefont
			{M.~J.}\ \bibnamefont {Stevens}}, \bibinfo {author} {\bibfnamefont
			{J.}~\bibnamefont {Tranchida}}, \bibinfo {author} {\bibfnamefont
			{C.}~\bibnamefont {Trott}},\ and\ \bibinfo {author} {\bibfnamefont {S.~J.}\
			\bibnamefont {Plimpton}},\ }\bibfield  {title} {\bibinfo {title} {Lammps - a
			flexible simulation tool for particle-based materials modeling at the atomic,
			meso, and continuum scales},\ }\href
	{https://doi.org/https://doi.org/10.1016/j.cpc.2021.108171} {\bibfield
		{journal} {\bibinfo  {journal} {Comput. Phys. Commun.}\ }\textbf {\bibinfo
			{volume} {271}},\ \bibinfo {pages} {108171} (\bibinfo {year}
		{2022})}\BibitemShut {NoStop}%
	\bibitem [{\citenamefont {Brown}\ and\ \citenamefont {Yamada}(2013)}]{lmp13}%
	\BibitemOpen
	\bibfield  {author} {\bibinfo {author} {\bibfnamefont {W.~M.}\ \bibnamefont
			{Brown}}\ and\ \bibinfo {author} {\bibfnamefont {M.}~\bibnamefont {Yamada}},\
	}\bibfield  {title} {\bibinfo {title} {Implementing molecular dynamics on
			hybrid high performance computers—three-body potentials},\ }\href
	{https://doi.org/https://doi.org/10.1016/j.cpc.2013.08.002} {\bibfield
		{journal} {\bibinfo  {journal} {Comput. Phys. Commun.}\ }\textbf {\bibinfo
			{volume} {184}},\ \bibinfo {pages} {2785} (\bibinfo {year}
		{2013})}\BibitemShut {NoStop}%
	\bibitem [{\citenamefont {Lyu}\ \emph {et~al.}(2021)\citenamefont {Lyu},
		\citenamefont {Qiao}, \citenamefont {Yao}, \citenamefont {Wang},
		\citenamefont {Morthomas}, \citenamefont {Fusco},\ and\ \citenamefont
		{Rodney}}]{lyu2021}%
	\BibitemOpen
	\bibfield  {author} {\bibinfo {author} {\bibfnamefont {G.-J.}\ \bibnamefont
			{Lyu}}, \bibinfo {author} {\bibfnamefont {J.-C.}\ \bibnamefont {Qiao}},
		\bibinfo {author} {\bibfnamefont {Y.}~\bibnamefont {Yao}}, \bibinfo {author}
		{\bibfnamefont {Y.-J.}\ \bibnamefont {Wang}}, \bibinfo {author}
		{\bibfnamefont {J.}~\bibnamefont {Morthomas}}, \bibinfo {author}
		{\bibfnamefont {C.}~\bibnamefont {Fusco}},\ and\ \bibinfo {author}
		{\bibfnamefont {D.}~\bibnamefont {Rodney}},\ }\bibfield  {title} {\bibinfo
		{title} {Microstructural effects on the dynamical relaxation of glasses and
			glass composites: {{A}} molecular dynamics study},\ }\href
	{https://doi.org/10.1016/j.actamat.2021.117293} {\bibfield  {journal}
		{\bibinfo  {journal} {Acta Mater.}\ }\textbf {\bibinfo {volume} {220}},\
		\bibinfo {pages} {117293} (\bibinfo {year} {2021})}\BibitemShut {NoStop}%
	\bibitem [{\citenamefont {Yu}\ \emph {et~al.}(2017)\citenamefont {Yu},
		\citenamefont {Richert},\ and\ \citenamefont {Samwer}}]{yu2017}%
	\BibitemOpen
	\bibfield  {author} {\bibinfo {author} {\bibfnamefont {H.-B.}\ \bibnamefont
			{Yu}}, \bibinfo {author} {\bibfnamefont {R.}~\bibnamefont {Richert}},\ and\
		\bibinfo {author} {\bibfnamefont {K.}~\bibnamefont {Samwer}},\ }\bibfield
	{title} {\bibinfo {title} {Structural rearrangements governing
			johari-goldstein relaxations in metallic glasses},\ }\href
	{https://doi.org/10.1126/sciadv.1701577} {\bibfield  {journal} {\bibinfo
			{journal} {Sci. Adv.}\ }\textbf {\bibinfo {volume} {3}},\ \bibinfo {pages}
		{e1701577} (\bibinfo {year} {2017})}\BibitemShut {NoStop}%
	\bibitem [{\citenamefont {Zella}\ \emph {et~al.}(2022)\citenamefont {Zella},
		\citenamefont {Moon}, \citenamefont {Keffer},\ and\ \citenamefont
		{Egami}}]{Takeshi2022}%
	\BibitemOpen
	\bibfield  {author} {\bibinfo {author} {\bibfnamefont {L.}~\bibnamefont
			{Zella}}, \bibinfo {author} {\bibfnamefont {J.}~\bibnamefont {Moon}},
		\bibinfo {author} {\bibfnamefont {D.}~\bibnamefont {Keffer}},\ and\ \bibinfo
		{author} {\bibfnamefont {T.}~\bibnamefont {Egami}},\ }\bibfield  {title}
	{\bibinfo {title} {Transient nature of fast relaxation in metallic glass},\
	}\href {https://doi.org/10.1016/j.actamat.2022.118254} {\bibfield  {journal}
		{\bibinfo  {journal} {Acta Mater.}\ }\textbf {\bibinfo {volume} {239}},\
		\bibinfo {pages} {118254} (\bibinfo {year} {2022})}\BibitemShut {NoStop}%
	\bibitem [{\citenamefont {Douglas}\ \emph {et~al.}(2024)\citenamefont
		{Douglas}, \citenamefont {Yuan}, \citenamefont {Zhang}, \citenamefont
		{Zhang},\ and\ \citenamefont {Xu}}]{Douglas2024}%
	\BibitemOpen
	\bibfield  {author} {\bibinfo {author} {\bibfnamefont {J.~F.}\ \bibnamefont
			{Douglas}}, \bibinfo {author} {\bibfnamefont {Q.-L.}\ \bibnamefont {Yuan}},
		\bibinfo {author} {\bibfnamefont {J.}~\bibnamefont {Zhang}}, \bibinfo
		{author} {\bibfnamefont {H.}~\bibnamefont {Zhang}},\ and\ \bibinfo {author}
		{\bibfnamefont {W.-S.}\ \bibnamefont {Xu}},\ }\bibfield  {title} {\bibinfo
		{title} {A dynamical system approach to relaxation in glass-forming
			liquids},\ }\href {https://doi.org/10.1039/D4SM00976B} {\bibfield  {journal}
		{\bibinfo  {journal} {Soft Matter}\ }\textbf {\bibinfo {volume} {20}},\
		\bibinfo {pages} {9140} (\bibinfo {year} {2024})}\BibitemShut {NoStop}%
	\bibitem [{\citenamefont {Ngai}(2023)}]{ngai2023}%
	\BibitemOpen
	\bibfield  {author} {\bibinfo {author} {\bibfnamefont {K.}~\bibnamefont
			{Ngai}},\ }\bibfield  {title} {\bibinfo {title} {Universal properties of
			relaxation and diffusion in complex materials: {{Originating}} from
			fundamental physics with rich applications},\ }\href
	{https://doi.org/10.1016/j.pmatsci.2023.101130} {\bibfield  {journal}
		{\bibinfo  {journal} {Prog. Mater Sci.}\ }\textbf {\bibinfo {volume} {139}},\
		\bibinfo {pages} {101130} (\bibinfo {year} {2023})}\BibitemShut {NoStop}%
	\bibitem [{\citenamefont {Zhang}\ \emph {et~al.}(2016)\citenamefont {Zhang},
		\citenamefont {Ashcraft}, \citenamefont {Mendelev}, \citenamefont {Wang},\
		and\ \citenamefont {Kelton}}]{Zhang16-NiNb}%
	\BibitemOpen
	\bibfield  {author} {\bibinfo {author} {\bibfnamefont {Y.}~\bibnamefont
			{Zhang}}, \bibinfo {author} {\bibfnamefont {R.}~\bibnamefont {Ashcraft}},
		\bibinfo {author} {\bibfnamefont {M.}~\bibnamefont {Mendelev}}, \bibinfo
		{author} {\bibfnamefont {C.~Z.}\ \bibnamefont {Wang}},\ and\ \bibinfo
		{author} {\bibfnamefont {K.~F.}\ \bibnamefont {Kelton}},\ }\bibfield  {title}
	{\bibinfo {title} {{Experimental and molecular dynamics simulation study of
				structure of liquid and amorphous Ni\textsubscript{62}Nb\textsubscript{38}
				alloy}},\ }\href {https://doi.org/10.1063/1.4968212} {\bibfield  {journal}
		{\bibinfo  {journal} {J. Chem. Phys.}\ }\textbf {\bibinfo {volume} {145}},\
		\bibinfo {pages} {204505} (\bibinfo {year} {2016})}\BibitemShut {NoStop}%
	\bibitem [{\citenamefont {Shiraishi}\ \emph {et~al.}(2023)\citenamefont
		{Shiraishi}, \citenamefont {Mizuno},\ and\ \citenamefont {Ikeda}}]{pel1}%
	\BibitemOpen
	\bibfield  {author} {\bibinfo {author} {\bibfnamefont {K.}~\bibnamefont
			{Shiraishi}}, \bibinfo {author} {\bibfnamefont {H.}~\bibnamefont {Mizuno}},\
		and\ \bibinfo {author} {\bibfnamefont {A.}~\bibnamefont {Ikeda}},\ }\bibfield
	{title} {\bibinfo {title} {Johari-goldstein $\beta$ relaxation in glassy
			dynamics originates from two-scale energy landscape},\ }\href
	{https://doi.org/10.1073/pnas.2215153120} {\bibfield  {journal} {\bibinfo
			{journal} {Proc. Natl. Acad. Sci.}\ }\textbf {\bibinfo {volume} {120}},\
		\bibinfo {pages} {e2215153120} (\bibinfo {year} {2023})}\BibitemShut
	{NoStop}%
	\bibitem [{\citenamefont {Ding}\ \emph {et~al.}(2016)\citenamefont {Ding},
		\citenamefont {Cheng}, \citenamefont {Sheng}, \citenamefont {Asta},
		\citenamefont {Ritchie},\ and\ \citenamefont {Ma}}]{dingnc2016}%
	\BibitemOpen
	\bibfield  {author} {\bibinfo {author} {\bibfnamefont {J.}~\bibnamefont
			{Ding}}, \bibinfo {author} {\bibfnamefont {Y.~Q.}\ \bibnamefont {Cheng}},
		\bibinfo {author} {\bibfnamefont {H.}~\bibnamefont {Sheng}}, \bibinfo
		{author} {\bibfnamefont {M.}~\bibnamefont {Asta}}, \bibinfo {author}
		{\bibfnamefont {R.~O.}\ \bibnamefont {Ritchie}},\ and\ \bibinfo {author}
		{\bibfnamefont {E.}~\bibnamefont {Ma}},\ }\bibfield  {title} {\bibinfo
		{title} {Universal structural parameter to quantitatively predict metallic
			glass properties},\ }\href {https://doi.org/10.1038/ncomms13733} {\bibfield
		{journal} {\bibinfo  {journal} {Nat. Commun.}\ }\textbf {\bibinfo {volume}
			{7}},\ \bibinfo {pages} {13733} (\bibinfo {year} {2016})}\BibitemShut
	{NoStop}%
	\bibitem [{\citenamefont {Zella}\ \emph {et~al.}(2024)\citenamefont {Zella},
		\citenamefont {Moon},\ and\ \citenamefont {Egami}}]{Takeshi2024}%
	\BibitemOpen
	\bibfield  {author} {\bibinfo {author} {\bibfnamefont {L.}~\bibnamefont
			{Zella}}, \bibinfo {author} {\bibfnamefont {J.}~\bibnamefont {Moon}},\ and\
		\bibinfo {author} {\bibfnamefont {T.}~\bibnamefont {Egami}},\ }\bibfield
	{title} {\bibinfo {title} {Ripples in the bottom of the potential energy
			landscape of metallic glass},\ }\href
	{https://doi.org/10.1038/s41467-024-45640-1} {\bibfield  {journal} {\bibinfo
			{journal} {Nat. Commun.}\ }\textbf {\bibinfo {volume} {15}},\ \bibinfo
		{pages} {1358} (\bibinfo {year} {2024})}\BibitemShut {NoStop}%
	\bibitem [{\citenamefont {Fan}\ \emph {et~al.}(2017)\citenamefont {Fan},
		\citenamefont {Iwashita},\ and\ \citenamefont {Egami}}]{fan2017}%
	\BibitemOpen
	\bibfield  {author} {\bibinfo {author} {\bibfnamefont {Y.}~\bibnamefont
			{Fan}}, \bibinfo {author} {\bibfnamefont {T.}~\bibnamefont {Iwashita}},\ and\
		\bibinfo {author} {\bibfnamefont {T.}~\bibnamefont {Egami}},\ }\bibfield
	{title} {\bibinfo {title} {Energy landscape-driven non-equilibrium evolution
			of inherent structure in disordered material},\ }\href
	{https://doi.org/10.1038/ncomms15417} {\bibfield  {journal} {\bibinfo
			{journal} {Nat. Commun.}\ }\textbf {\bibinfo {volume} {8}},\ \bibinfo {pages}
		{15417} (\bibinfo {year} {2017})}\BibitemShut {NoStop}%
	\bibitem [{\citenamefont {Hara}\ \emph {et~al.}(2025)\citenamefont {Hara},
		\citenamefont {Matsuoka}, \citenamefont {Ebata}, \citenamefont {Mizuno},\
		and\ \citenamefont {Ikeda}}]{haraL2025a}%
	\BibitemOpen
	\bibfield  {author} {\bibinfo {author} {\bibfnamefont {Y.}~\bibnamefont
			{Hara}}, \bibinfo {author} {\bibfnamefont {R.}~\bibnamefont {Matsuoka}},
		\bibinfo {author} {\bibfnamefont {H.}~\bibnamefont {Ebata}}, \bibinfo
		{author} {\bibfnamefont {D.}~\bibnamefont {Mizuno}},\ and\ \bibinfo {author}
		{\bibfnamefont {A.}~\bibnamefont {Ikeda}},\ }\bibfield  {title} {\bibinfo
		{title} {A link between anomalous viscous loss and the boson peak in soft
			jammed solids},\ }\href {https://doi.org/10.1038/s41567-024-02722-7}
	{\bibfield  {journal} {\bibinfo  {journal} {Nat. Phys.}\ }\textbf {\bibinfo
			{volume} {21}},\ \bibinfo {pages} {262} (\bibinfo {year} {2025})}\BibitemShut
	{NoStop}%
\end{thebibliography}
\end{document}